\documentclass[nofootinbib, preprintnumbers, superscriptaddress,notitlepage]{revtex4-1}
\usepackage{amsmath,amssymb,amscd,simplewick,braket,verbatim}
\usepackage{listings}
\usepackage{dsfont}
\usepackage{slashed}
\usepackage{color}
\usepackage{ulem}
\usepackage{float}

\usepackage{graphicx}
\usepackage{epstopdf}
\usepackage{subfigure}
\usepackage{epsfig}

\usepackage{xcolor}
\usepackage[colorlinks=true,
            linkcolor=blue,
            urlcolor=blue,
            citecolor=green,          
            bookmarks=true,
            bookmarksnumbered=true,
            breaklinks=true,
            pdfpagemode=Fullscreen,
            pdfstartview=FitBH]{hyperref}

\hypersetup{pdfauthor = {Shao-Feng Ge},
	     pdftitle = {}, % Insert title here!!!
	     pdfsubject = {}, % Insert subject here!!!
             pdfkeywords = {}, % Insert keywords here!!!
	     pdfcreator = {LaTeX with hyperref package},
	     pdfproducer = {dvips + ps2pdf} }

\definecolor{gesfpurple}{rgb}{0.47,0.19,0.42}

\definecolor{gesflanse}{rgb}{0.00,0.50,0.50}

\definecolor{gesfblue}{rgb}{0.08,0.42,0.76}

\definecolor{gesfred}{rgb}{1,0,0}

\definecolor{gesfwhite}{rgb}{1,1,1}

\definecolor{gesfblack}{rgb}{0,0,0}

\newcommand{\tr}{\mbox{Tr}}
\definecolor{OliveGreen}{cmyk}{0.64,0,0.95,0.40}

\newcommand{\gsec}[1]{{\hypersetup{linkcolor=red}Sec.~\ref{#1}\hypersetup{linkcolor=blue}}}

\newcommand{\geqn}[1]{\hypersetup{linkcolor=blue}(\ref{#1})\hypersetup{linkcolor=blue}}
\newcommand{\gfig}[1]{{\hypersetup{linkcolor=violet}Fig.~\ref{#1}\hypersetup{linkcolor=blue}}}

\newcommand{\fp}{\slashed p}

\begin{document}
\fontsize{12pt}{14pt}\selectfont

\title{\Large Solar Active-Sterile Neutrino Conversion with Atomic Effects \\ at Dark Matter Direct Detection Experiments}
\author{Shao-Feng Ge}
\email{gesf@sjtu.edu.cn}
\affiliation{Tsung-Dao Lee Institute (TDLI) \& School of Physics and Astronomy (SPA),
Shanghai Jiao Tong University (SJTU), Shanghai 200240, China}
\affiliation{Key Laboratory for Particle Astrophysics and Cosmology (MOE) \& Shanghai Key Laboratory for Particle Physics and Cosmology, Shanghai Jiao Tong University, Shanghai 200240, China}

\author{Pedro Pasquini}
\email{ppasquini@sjtu.edu.cn}
\affiliation{Tsung-Dao Lee Institute (TDLI) \& School of Physics and Astronomy (SPA),
Shanghai Jiao Tong University (SJTU), Shanghai 200240, China}
\affiliation{Key Laboratory for Particle Astrophysics and Cosmology (MOE) \& Shanghai Key Laboratory for Particle Physics and Cosmology, Shanghai Jiao Tong University, Shanghai 200240, China}

\author{Jie Sheng}
\email{shengjie04@sjtu.edu.cn}
\affiliation{Tsung-Dao Lee Institute (TDLI) \& School of Physics and Astronomy (SPA),
Shanghai Jiao Tong University (SJTU), Shanghai 200240, China}
\affiliation{Key Laboratory for Particle Astrophysics and Cosmology (MOE) \& Shanghai Key Laboratory for Particle Physics and Cosmology, Shanghai Jiao Tong University, Shanghai 200240, China}

\begin{abstract}

The recent XENON1T excess can be explained by the solar
active-sterile neutrino conversion with bound electrons
via light mediator. Nevertheless, the atomic effects are usually
omitted in the solar neutrino explanations.
We systematically establish a second quantization formalism 
for both bound and ionized electrons to account for the
atomic effects. This formalism is of great generality to 
incorporate various interactions
for both neutrino and dark matter scatterings.
Our calculation shows that the change in the cross section due 
to atomic effects can have important impact on the
differential cross section. It is necessary
to include atomic effects in the low-energy
electron recoil signal at dark matter direct detection
experiments even for energetic solar neutrinos.
With the best-fit values to the XENON1T data, we also
project the event rate at PandaX-4T, XENONnT, and LZ
experiments. 

\end{abstract}

\maketitle

\section{Introduction}

The particle nature and properties of dark matter (DM)
can be probed by the direct detection experiments, typically utilizing nuclear or electron recoils \cite{Schumann:2019eaa}.
Recently, the XENON1T experiment found an excess in the electron recoil spectrum around $(2 \sim 3)$\,keV \cite{Aprile:2020tmw} which was independently checked by the PandaX-II
experiment \cite{PandaX-II:2020udv}. Although
solar neutrinos can also scatter with electron via the Standard Model (SM) weak interactions, this signal can contribute only $0.52\%$ of the total events as background and more 
importantly the electron recoil spectrum is quite flat in the observed keV range \cite{Jeong:2021ivd}. The excess might indicate new physics beyond the SM (BSM) if not the tritium background \cite{Robinson:2020gfu}. Possible explanations
include axion or axion-like particles
\cite{Takahashi:2020bpq,DiLuzio:2020jjp,Gao:2020wer,
Dent:2020jhf,OHare:2020wum,Dessert:2020vxy,Li:2020naa,
Athron:2020maw,Han:2020dwo},
elastic \cite{Chen:2020gcl,Cao:2020bwd,
Su:2020zny,Paz:2020pbc,Nakayama:2020ikz,Jho:2020sku,
Zu:2020idx,DelleRose:2020pbh,Alhazmi:2020fju,
Ko:2020gdg,Chigusa:2020bgq} 
and inelastic \cite{Harigaya:2020ckz,Baryakhtar:2020rwy,
An:2020tcg,Lee:2020wmh,Bramante:2020zos,An:2020bxd,
Bloch:2020uzh,Dutta:2021wbn,He:2020sat,Guo:2020oum,
Borah:2020jzi,Dror:2020czw} 
scattering between 
DM and electron, the Migdal effect of DM scattering with
nuclei \cite{Dey:2020sai}, and DM decay 
\cite{Choi:2020udy,Bell:2020bes,Du:2020ybt,
Du:2020ybt,Hryczuk:2020jhi}.
Another possibility is sterile neutrino as DM in our galaxy.
Either the sterile neutrino DM inelastically scatters with
electron into active neutrino and releases its mass as energy
in electron recoil \cite{Shakeri:2020wvk,Xue:2020cnw} or sterile
neutrino decays inside the detector to produce a photon signal
\cite{Khruschov:2020cnf}.

In addition, light neutrinos can also provide an explanation.
Especially, the Sun is a natural source of energetic neutrino
fluxes around Earth. While the SM neutrino interactions can only produce a flat event spectrum, a low energy peak
arises if there is a BSM interaction with light mediator. Possible
realizations include magnetic dipole moment,
charge radius and anapole 
\cite{Bell:2005kz,Bell:2006wi,Miranda:2020kwy,Babu:2020ivd,
Brdar:2021xll,Ye:2021zso,Ni:2021mwa,Yue:2021vjg,
PandaX-II:2020udv,AristizabalSierra:2020zod},
all with the massless photon as mediator.
In addition, light scalar \cite{Khan:2020vaf,Boehm:2020ltd,Alikhanov:2021dhb}
and vector \cite{Chen:2021uuw,Ibe:2020dly,Alikhanov:2021dhb, Khan:2020vaf,
Boehm:2020ltd,Bally:2020yid,AristizabalSierra:2020edu,
Lindner:2020kko,Karmakar:2020rbi} mediators can also achieve the
same purpose. It is interesting to observe that only
scalar and vector mediators have been used to explain the
low energy peak while the pseudo-scalar one was claimed to be
incapable of achieving the goal \cite{Boehm:2020ltd}.
However, if neutrino scattering with electron into a massive
sterile neutrino in the final state, a light pseudo-scalar
mediator can also produce a peak in the low recoil energy
\cite{Ge:2020jfn}. The $\mathcal O(100)$ keV sterile neutrino can also be applied with
dipole magnetic moment interactions to explain the excess \cite{Shoemaker:2020kji}.

For an $\mathcal O($keV) electron recoil energy, the momentum transfer
is comparable with the atomic energy of a heavy element such
as Xe. The electrons inside atom can no longer be treated as
a free particle. Instead, one should use quantum wave functions
to describe the electron distribution.
It is the electron cloud rather than a single electron
particle that participates in the scattering process. This has been extensively
explored for the DM scattering with electron in recent years
\cite{Essig:2011nj,Essig:2012yx,Essig:2015cda,
Roberts:2016xfw,Essig:2017kqs,Catena:2019gfa}. The
effect of the initial- and final-state electron wave
functions can be summarized into a $K$-factor
\cite{Catena:2019gfa} or sometimes $f$-factor
\cite{Essig:2011nj} as function of the momentum
transfer $\bf q$. The size of atomic effect is typically
$\mathcal O(1)$ and hence not negligible.

Being usually unnoticed, the atomic effect with electron
cloud has also been considered in the
study of neutrino electromagnetic properties.
Since the massless photon is exactly a light mediator,
the momentum transfer in neutrino scattering
with photon mediation is intrinsically suppressed in the
same way as the low energy electron recoil at DM
detectors. So the atomic effect cannot be neglected in
the study of neutrino electromagnetic properties. For
example, the ionization effect due to the neutrino
scattering with bound electrons is obtained by
considering the binding energy and the initial wave
function in light atoms \cite{Gounaris:2001jp,
Gounaris:2004ji}. Later, the effect of atomic potential
on the final-state electron is also taken into
consideration when calculating the cross section 
\cite{Voloshin:2010vm,Kouzakov:2010tx,Kouzakov:2011vx}.
Although the atomic effect was claimed to be small
\cite{Kouzakov:2011vx}, recent studies
realize that it is actually not
negligible \cite{Chen:2013iud,Chen:2013lba}.
Even spin effects have been studied very lately
\cite{Whittingham:2021mdw}. 

In this work, we study the neutrino scattering with
bound electrons into a massive sterile neutrino to
explain the observed electron recoil peak at $(2 \sim 3)$\,keV
and make projections for the future DM experiments.
In \gsec{sec:AtomicEffect},
we first summarize the formalism of calculating
the neutrino scattering with both free and bound electrons
in the language of second quantization.
Especially, the bound electron is directly quantized using
annihilation and creation operators without involving
the inappropriate plane waves or equivalently momentum eigenstates.
\gsec{sec:CrossSections} shows the cross 
section of neutrino electron scattering including
the atomic effect and 
compare it with the free electron scattering case.
The results are further used
in \gsec{sec:Experiments} to fit the observed data
with $\chi^2$ minimization to constrain the parameter space
and the prospect of detecting such signal at future
experiments is projected in \gsec{Section:Predictions}.
Finally we summarize and conclude in \gsec{sec:summary}.

\section{Neutrino-Electron Scattering With Atomic Effects}
\label{sec:AtomicEffect}

As mentioned above, the atomic effects need to be
considered in order to make a precise study of the
electron recoil signal from DM or neutrino scattering.
The initial-state electron is bound inside the atom
instead of being free. In contrast, the final-state
electron is knocked out of the atom and becomes ionized
leaving a recoil signal in the detector. With a recoil
energy around $(2 \sim 3)$\,keV, the ionized electron
is not completely free but is subject to the atomic
Coulomb potential. One needs to consider the atomic
effect for both the initial- and final-state electrons.
We first try to establish a unified second quantization
description of the bound and ionized states in 
\gsec{sec:2ndQ} and then use it to calculate the
atomic $K$-factor in \gsec{sec:atomic}.
This formalism of second quantization can accommodate
general interactions and apply not just for neutrino but
also DM scatterings. 

\subsection{Second Quantization of Bound and Ionized Electron States}
\label{sec:2ndQ}

An electron trapped in the Coulomb potential is no
longer a free particle and hence cannot be described by
plane wave with fixed momentum. Instead, the conserved
variable is the energy eigenvalue including both kinetic
energy and Coulomb potential. To be concrete, the bound
electron field $e_B(x)$ is in general a function
of spatial coordinates \cite{Weinberg:1995mt},
\begin{eqnarray}
  e_B(x)
=
  \sum_{nlm} a_{nlm} e^{-i E_{nl} t} \psi_{nlm} ({\bf x}),
\label{eq:eBound}
\end{eqnarray}
where $a_{nlm}$ ($a^\dagger_{nlm}$) is the creation
(annihilation) operator for the bound state with
principal ($n$), angular ($l$), and magnetic ($m$)
quantum numbers. Since positron never enters our
discussion, we can safely omit $b^\dagger_{nlm}$ in
\geqn{eq:eBound} from the beginning. When acting on the
vaccum state $|0\rangle$, the creation operator $a^\dagger_{nlm}$
gives the bound state $|nlm\rangle \equiv
a^\dagger_{nlm} | 0 \rangle$. There is no need to
involve the $\sqrt{2 E}$ prefactor that is necessary for
a relativistic particle to keep Lorentz covariance
but reduces to a constant $\sqrt{2 m}$ and hence
can be combined into normalization for a
non-relativistic particle. Similarly, we follow the
convention of second quantization 
to define the anti-commutation relations,
$\{a_{nlm}, a^\dagger_{n'l'm'}\} = \delta_{nn'}
\delta_{ll'} \delta_{mm'}$ and $\{a_{nlm}, a_{n'l'm'}\}
= \{a^\dagger_{nlm}, a^\dagger_{n'l'm'}\} = 0$.
With dimensionless discrete $\delta$ functions, the
creation and annihilation operators are also
dimensionless, $[a_{nlm}] = [a^\dagger_{nlm}] = 0$, and
the bound electron field has the same dimension as its
wave function, $[e_B(x)] = [\psi_{nlm}({\bf x})]$.
The normalization condition of the wave function,
$\int \psi^\dagger_{nlm}({\bf x}) \psi_{n'l'm'}({\bf x})
d^3 {\bf x} = \delta_{nn'} \delta_{ll'} \delta_{mm'}$,
fixes the field dimension to be $[e_B(x)] = 3/2$ which
is consistent with quantum field theory (QFT). 
Although the momentum integration is replaced by a
summation $\sum_{nlm}$, the second quantized field
$e_B(x)$ is still a linear combination of energy and
angular momentum eigenstates. However, the second
quantized field $e_B(x)$ in \geqn{eq:eBound} is
different from the usual formalism of a free particle in
quantum field theory \cite{Peskin:1995ev} with the
evolution phase $e^{- i E_{nl} t}$ containing only the
energy eigenvalue and time dependence while the spatial
dependence in the electron wave function
$\psi_{nlm}({\bf x})$ cannot be factorized out as a
simple complex phase.

For an ionized electron that is still under the
influence of the atomic Coulomb potential, its energy
is continuously distributed. The
corresponding ionized electron field contains an integration
over the asymptotic momentum,
\begin{gather}
   e_I (x)
=
  \sum_{lm} \int \frac{|{\bf p}|^2 d |{\bf p}|}{(2 \pi)^3}
  a_{T_r lm}\psi_{T_r lm}({\bf x}) e^{-i T_r t},
\label{eq:eIonized}
\end{gather}
without involving a principal quantum number. The
asymptotic kinetic energy $T_r \equiv |{\bf p}|^2 / 2
m_e$ is the one that we can experimentally measure.
Comparing with the bound case in \geqn{eq:eBound}, the
only difference is that the energy eigenvalue changes
from a discrete principal quantum number $n$ to a
continuous $T_r$ or equivalently the asymptotic
momentum $|{\bf p}|$. Similar to the summation over
discrete variables, the asymptotic momentum is
integrated. In other words, the second quantized field
$e_I(x)$ is also a linear combination of energy and
angular momentum eigenstates. With one-to-one
correspondence between the asymptotic energy $T_r$ and
the asymptotic momentum $|{\bf p}|$, the latter is also
a well defined physical quantity. The ionized electron
behaves essentially as a free particle when $|{\bf x}|
\rightarrow \infty$ and its wave function reduces to a
plane wave \cite{Bethe:1957ncq}. The direction of the
asymptotic momentum can also be used to label an ionized
electron in the similar way as its magnitude. The
Fourier transformation for the asymptotic state at
infinity distance ($|{\bf x}| \rightarrow \infty$) is,
$\psi({\bf x}) = \int \frac {d^3 {\bf p}}{(2 \pi)^3}
\psi({\bf p}) e^{- i {\bf p} \cdot {\bf x}}$.
However, the DM direct detection experiments cannot tell
the directional information and are sensitive to only
the magnitude $|{\bf p}|$ or the recoil energy $T_r$.
Namely, the relevant wave function is
$\psi_{T_r}({\bf x}) \equiv \int d \Omega_{\bf p}
\psi({\bf p}) e^{- i {\bf p} \cdot {\bf x}}$ and the
remaining phase space integration
is exactly the $|{\bf p}|^2 d |{\bf p}|$ used in \geqn{eq:eIonized}, 
$\psi({\bf x}) = \int \frac {|{\bf
p}|^2 d |{\bf p}|}{(2 \pi)^3} \psi_{T_r}({\bf x})$.
The original anti-commutation relation 
$
  \{ a_{{\bf p} lm}, a^\dagger_{{\bf p}' l'm'}\} 
= 
  (2 \pi)^3 
  \delta_{ll'}
  \delta_{mm'} 
  \delta^{(3)}({\bf p} - {\bf p}')
$
reduces to 
$
  \{ a_{T_r lm}, a^\dagger_{T'_r l'm'}\} 
=
  (2 \pi)^3 \delta_{ll'} 
  \delta_{mm'} 
  \delta(|{\bf p}| - |{\bf p}'|) / |{\bf p}|^2
$ 
after integrating away the angular information of
the asymptotic momentum ${\bf p}$. Correspondingly, the
operator dimension is, $[a_{T_r lm}] = - 3/2$ and the
wave function $\psi_{T_r lm}({\bf x})$ for ionized state
is dimensionless. This is consistent with the wave function normalization, 
$
  \int 
    \psi^\dagger_{T_r lm}({\bf x}) 
    \psi_{T'_r l'm'}({\bf x}) 
  d^3 {\bf x}
=
  (2 \pi)^3 
  \delta_{ll'} 
  \delta_{mm'} 
  \delta(|{\bf p}| - |{\bf p}'|) / |{\bf p}|^2
$ \cite{Essig:2015cda}. Using the non-relativistic dispersion relation, $T_r = |{\bf p}|^2 / 2 m_e$, one can prove that, 
$
\left[ \delta(|{\bf p}| - |{\bf p}'|) / |{\bf p}|^2 \right]
\times 
  |{\bf p}|^2 d |{\bf p}| 
= 
  \delta (T_r - T'_r) d T_r
$
and the normalization condition becomes,
$
  \int 
    \psi^\dagger_{T_r lm}({\bf x}) 
    \psi_{T'_r l'm'}({\bf x}) 
  d^3 {\bf x}
=
  (2 \pi)^3 
  \delta_{ll'} 
  \delta_{mm'} 
  \delta(T_r - T'_r)
$ \cite{Bethe:1957ncq} to make everything consistent.
Similar to the bound state, an
ionized state $| T_r lm \rangle \equiv a^\dagger_{T_r lm} | 0 \rangle$ is created from vacuum by $a^\dagger_{T_r lm}$ without involving the energy prefactor $\sqrt{2 E}$.

The wave function can be constructed from the field,
$\braket{0|e_B(x)|nlm} = \psi_{nlm} ({\bf x}) e^{- i E_{nl} t}$ for the bound state and
$
  \braket{0|e_I(x)|T_r lm} 
= 
  \psi_{T_r lm} ({\bf x}) 
  e^{- i T_r t}
$ for the ionized one. In addition to the spatial distribution, the wave function should also contain the spin information. The spinor wave function
$\psi_N({\bf x})$ is a solution of the covariant Dirac equation in the presence of electromagnetic field,
$\left[ i \slashed \partial - m - e \slashed A  \right] \psi_N(x) = 0$, where $N \equiv (nlm)$ and
$N \equiv (T_r lm)$ for the bound and ionized states,
respectively. For a stationary energy eigenstate,
$\psi_N(x) = e^{- i E_N t} \psi_N ({\bf x})$,
the time dependence can be removed from the Dirac equation,
\begin{equation}
  \gamma^0
  [i{\bf \boldsymbol \gamma \cdot \boldsymbol \nabla} + m] \psi_N ({\bf x})
=
  [E_N - V(|\bf x|)] \psi_N ({\bf x}).
\end{equation}
For clarity, we have only kept the electric potential
$V(|{\bf x}|)$ in the atom as a central force.
In the non-relativistic limit, the spatial and spin parts
separate into \cite{Weinberg:1995mt},
\begin{equation}
  \psi_{N,s} (\textbf{x})
\approx
  \frac{1}{\sqrt{2}}
\left\lgroup
\begin{matrix}
  (1 + i {\boldsymbol \sigma \cdot \boldsymbol \nabla} / 2 m_e) f_{N,s} (\textbf{x}) \\
  (1 - i {\boldsymbol \sigma \cdot \boldsymbol \nabla} / 2 m_e) f_{N,s} (\textbf{x})
\end{matrix}
\right\rgroup,
\qquad
  f_{N,s} (\textbf{x})
\equiv
  \phi_N (\textbf{x}) \xi_s,
\label{DiracSolution1}
\end{equation}  
for spin $s$ and other quantum numbers $N$. The
information of spatial distribution and spin is
represented by the single-valued wave function
$\phi_N({\bf x})$ and the two-component spinor $\xi_s$,
respectively. The normalization condition can be
rewritten in terms of
$\phi_N({\bf x})$,
$
  \int d^3 {\bf x} 
    \psi^\dagger_{N,s}({\bf x}) 
    \psi_{N',s'}({\bf x})
\approx
  \int d^3 {\bf x} 
    \phi^*_N({\bf x}) 
    \phi_{N'}({\bf x})
=
  1
$. For a heavy element such as Xenon, the binding energy
is typically $\mathcal O(1 \sim 10)$\,keV. According to
the Virial theorem, the electron kinetic energy is of
the same size, which is roughly $0.1 \% \sim 1\%$ of the
electron mass, $m_e \approx 511$\,keV. In other words,
the effect of special relativity or spinor structure
represented by the gradient expansion is
roughly $0.1\% \sim 1\%$.

\subsection{Atomic Effects in Neutrino-Electron Scattering}
\label{sec:atomic}

With the second quantization formalism of the bound and ionized
states established, we can follow the usual QFT calculation of
transition rate and differential distributions.
For a general four-fermion coupling
$(\bar \nu \Gamma_\nu \nu_{s}) (\bar e \Gamma_e e)$ between
neutrino and electron, the transition matrix element
$\mathcal T$ is, 
\begin{equation}
  \mathcal{T}
\equiv
  \int d^4x d^4 y
  \langle T_rl'm'; {\bf p}_s| 
  \left[\overline \nu(x) \Gamma_\nu \nu_s(x) \phi^*(x)\right]
  \left[\overline e(y) \Gamma_e e(y) \phi(y)\right]|nlm; {\bf p}_\nu \rangle,
\end{equation}
with $\phi$ generally denoting a mediator with mass $M$
and ${\bf p}_\nu$ (${\bf p}_s$) is the solar (sterile)
neutrino momentum. For generality,
$
  \Gamma_\nu, \Gamma_e 
\equiv
  1, \gamma_5, \gamma^\mu, \gamma_5
  \gamma^\mu,
  \sigma^{\mu \nu}
$ denote all the possible Lorentz structures. 
Correspondingly, the mediator $\phi$ can be
a scalar (S), pseudo-scalar (P), vector (V),
axial-vector (A), or even tensor particle. The $\mathcal T$
matrix element should also contain coupling constants
which are omitted for simplicity. Note that the tensor
case is included just for completeness when illustrating
the atomic effects but not discussed when being applied
to the realistic case of sterile-active neutrino conversion.
This is because the tensor current is typically
mediated by a spin-2 particle such as the graviton that is beyond
the scope of this work. The other types with scalar, pseudo-scalar,
or vector mediator will be elaborated in the coming
\gsec{sec:CrossSections} and \gsec{sec:Experiments}.

When acting field operators on the initial 
$
  \ket{nlm, p_\nu} 
\equiv 
  a^\dagger_{nlm} 
  a^\dagger_{{\bf p}_\nu} 
  | 0 \rangle
$
and final 
$
  \ket{T_r lm, p_s} 
\equiv 
  a^\dagger_{T_r lm} 
  a^\dagger_{{\bf p}_s} 
  | 0 \rangle
$ states, the $a_{nlm}$ operator from $e_B(y$) annihilates a
bound state while the $a^\dagger_{T_r lm}$ operator from
$\bar e_I(y)$ creates an ionized state at position $\bf
y$. The $\mathcal T$ matrix becomes,
\begin{equation}
  \mathcal T
=
  \bar{u}_{\nu}\left(p_{\nu}\right) \Gamma_{\nu} 
  u_{s}\left(p_s\right)
  \frac{i}{q^2 - M^2}
  \int d^{4} y e^{-i q \cdot y}
  \overline \psi_{T_r l'm'}(\textbf{y}) \Gamma_e \psi_{nlm}(\textbf{y})
  e^{i \Delta E_{nl} t},
\label{eq:generalTmatrix}
\end{equation} 
after integrating away first the coordinate $x$ and the
momentum transfer $q$. The $x$ integration produces a
$\delta$-function of four momentum, $\delta^{(4)}(p_\nu - p_s - q)$, to
impose energy momentum conservation on the neutrino vertex,
$q = p_\nu - p_s$. From the bound and ionized electron
fields, one can extract an explicit energy dependence,
$\Delta E_{nl} \equiv T_r - E_{nl}$, as the
exponential factor $e^{i \Delta E_{nl} t}$.
This allows imposing energy conservation on the electron
vertex by integating away the time component $y^0$,
\begin{eqnarray}    
  \mathcal T
& = &
    \frac{i\bar{u}_{\nu}\left(p_{\nu}\right) \Gamma_{\nu} 
    u_{s}\left(p_s\right)}{q^2 - M^2}
    \int d^3 {\bf y} e^{i\textbf{q} \cdot \textbf{y}} 
    \overline \psi_{T_r l'm'}
    (\textbf{y}) \Gamma_e
    \psi_{nlm} (\textbf{y})
\nonumber
\\
& &
\hspace{35mm}
\times
    \delta
    \left(\Delta E_{nl} 
    - E_{\nu}
    + \sqrt{ m_s^2 + \left|\mathbf{p}_{\nu}\right|^{2}+|\mathbf{q}|^{2}
    -
    2\left|\mathbf{p}_{\nu} \| \mathbf{q}\right| \cos \theta_{q v}}\right).
\label{eq:Tenergydelta}
\end{eqnarray} 
Together with the $\delta^{(4)}(p_\nu - p_s - q)$ function
on the neutrino side, energy is conserved both locally and globally
while the momentum conservation only applies to the neutrino
vertex. This is the key difference between the calculations of
the scattering process with free or bound electron.
For free electron, the spatial dependence of their wave functions
can also factorize out as exponential factors and the integration
$\int d^4 y$ can impose both energy and momentum conservations.
The physical reason behind this is that a bound
electron does not have a definite momentum, especially in 
the coordinate representation. Consequently, we can only get
a product of the initial and final wave functions. 

Note that the 
$
  \delta_E
\equiv
  \delta
    (
      \Delta E_{nl} 
    -
      E_{\nu}
    +
      \sqrt{ 
        m_s^2 
      + 
        \left|\mathbf{p}_{\nu}\right|^{2}
      +
        |\mathbf{q}|^{2}
      -
        2
        \left|
          \mathbf{p}_{\nu} 
        \| 
          \mathbf{q}\right| 
        \cos \theta_{q v}
      }
    )
$ function in \geqn{eq:Tenergydelta} for energy
conservation can be moved outside of the $\int d^3 {\bf
y}$ integration since it only depends on energy and
momentum. The spatial integration with bound and ionized
wave functions, 
$
  A_e 
\equiv 
  \int d^3 {\bf y}
    e^{i\textbf{q} \cdot \textbf{y}}
    \overline \psi_{T_r l'm'} (\textbf{y}) 
    \Gamma_e 
    \psi_{nlm} (\textbf{y})
$,
is essentially the source
of the atomic $K$-factor. Nevertheless, the wave functions
$\psi_{nlm}$ and $\psi_{T_r l'm'}$ have spinor $\xi$ that
needs to be singled out in order to obtain the $\mathcal M$
matrix element.
For a scalar interaction ($\Gamma_e = 1$), the electron
amplitude $A_e$ divides into two parts,
\begin{eqnarray}
  A^S_e
\equiv
  \frac {\bar u(m_e) u(m_e)}{2 m_e}
    \int d^3 \textbf{y}
    e^{i\textbf{q} \cdot \textbf{y}}
    \phi_{T_r l'm'}^{*} (\textbf{y}) 
    \phi_{nlm} (\textbf{y}),
\end{eqnarray}
when expanded to the linear order of $i \boldsymbol \nabla f_{N,s}$.
The two-component spinor $\xi$ that is momentum independent
has been reexpressed, $\bar{u}(m_e) u(m_e) = 2m_e \xi^\dagger \xi$,
in terms of the electron spinor $u(m_e)$ at rest. The spinor and
spatial wave function then factorize into two parts,
$A^S_e \equiv [\bar u(m_e) u(m_e) / (2 m_e)] \times
f^{T_r l'm'}_{nlm}({\bf q})$ where
$f^{T_r l'm'}_{nlm}({\bf q})$ is the so-called atomic form factor
\cite{Essig:2011nj,Essig:2015cda,Catena:2019gfa},
\begin{equation}
  f^{T_r l'm'}_{nlm}({\bf q})
\equiv
    \int d^3 \textbf{y}
    e^{i\textbf{q} \cdot \textbf{y}}
    \phi_{T_r l'm'}^{*} (\textbf{y}) 
    \phi_{nlm} (\textbf{y}).
\label{formfactor-1}
\end{equation}

The factorization of the scattering amplitude into two
parts, one for the spinor and the other for the atomic
form factor, is a generic feature when expanding to the
leading order. For all the possible electron bilinears,
\begin{eqnarray}
  A^{S,P,V,A,T}_e
\equiv
  \bar u(m_e)
\left\{
  \frac 1 {2 m_e},
  \frac {\gamma_0 {\bf q} \cdot \boldsymbol \gamma \gamma^5} {4 m^2_e},
  \frac {\gamma^0} {2 m_e},
  \frac {\gamma^5 \gamma^i} {2 m_e},
  \frac {[\gamma^i, \gamma^j]} {2 m_e}
\right\}
  u(m_e)
  f^{T_r l'm'}_{nlm}({\bf q}).
\end{eqnarray}
The key feature here is that the initial- and
final-state electrons are not momentum eigenstates or
free-particle solution of the Dirac equations. Instead,
the bound and ionized spinors are originally a function
of the spatial coordinates \geqn{DiracSolution1}. There
is no need to involve initial and final momentum for
electrons at all. The only momentum that can appear is
the momentum transfer ${\bf q}$. Especially for the
pseudo-scalar case, the electron spin or equivalently
$\gamma^i$ with a spatial index is relevant. However,
only the inner product ${\bf q} \cdot \boldsymbol \gamma$
can appear as a combination to keep the scalar nature of this
vertex. This factorization is derived with second
quantization for the bound and ionized electron wave
functions in a systematic way.

With $A_e$ factorization, the $\mathcal T$ matrix element \geqn{eq:Tenergydelta} is composed of several contributions, 
$
  \mathcal T 
\equiv 
  \mathcal M f^{T_r l'm'}_{nlm}({\bf q}) 
  \delta_E
$,
where $\mathcal M$ is the scattering matrix element,
\begin{eqnarray}
  \mathcal M
\equiv
  \frac{i\bar{u}_{\nu}\left(p_{\nu}\right) \Gamma_{\nu} 
  u_{s}\left(p_s\right)}{q^2 - M^2}
  \bar u(m_e)
\left\{
  \frac 1 {2 m_e},
  \frac {\gamma_0 {\bf q} \cdot \boldsymbol \gamma \gamma^5} {4 m^2_e},
  \frac {\gamma^0} {2 m_e},
  \frac {\gamma^5 \gamma^i} {2 m_e},
  \frac {[\gamma^i, \gamma^j]} {2 m_e}
\right\}
  u(m_e).
\label{eq:M}
\end{eqnarray}
Note that $\mathcal M$ is
a function of the incoming neutrino momentum $p_\nu$ and the
momentum transfer ${\bf q}$ while the sterile neutrino momentum
$p_s = p_\nu - q$ is a dependent variable.

Following the derivations
in the Section 4.5 of \cite{Peskin:1995ev}, the scattering
cross section with a bound electron can be also expressed in
terms of $\mathcal M$,
\begin{eqnarray}
  \sigma_{nl}
\equiv
  \frac 1 {2 l + 1}
  \int \frac{|{\bf p}|^2 d |{\bf p}|}{(2 \pi)^3}
  \int \frac{d^3 {\bf q}}{(2\pi)^3 2 E_s}
  \frac{1}{8 m_e^2 E_\nu}
  \overline{|\mathcal M|^2} \,
  (2 \pi) \, \delta_E
  \sum_{m,\,l'm'}
  \left| f^{T_r l'm'}_{nlm}({\bf q}) \right|^2,
\label{eq:sigma6}
\end{eqnarray}
for a single electron target. The prefactor $1/(2l+1)$ accounts
for the degenerate electrons with $m = -l, -l+1, \cdots, l-1, l$
while $\overline{|\mathcal M|^2} \equiv |\mathcal M|^2 /
2$ is averaged over the electron spin. For the
high energy solar neutrino that contains mainly the
left-handed neutrinos, there is no need to average over
the neutrino spin. Since the asymptotic kinetic energy
$T_r = |{\bf p}|^2 / 2 m_e$ is the one that can be
experimentally measured, it is much more convenient to
express the phase space integration 
$
  |{\bf p}|^2 d |{\bf p}| 
=
  \frac 1 2 (2 m_e T_r)^{3/2} 
  d \ln T_r$
in terms of $T_r$. The original phase space integration
for the sterile neutrino momentum ${\bf p}_s$ has also
been replaced by the momentum transfer ${\bf q}$ due to
momentum conservation of the neutrino vertex. In
addition, the combined neutrino and atom system has
rotational invariance around the incoming neutrino momentum ${\bf p}_\nu$. So the azimuthal angle of ${\bf
q}$ can be integrated away to give $2 \pi$. The zenith angle $\theta_{q\nu}$ integration is reduced by
the $\delta_E$ to 
$
  \int \delta_E d 
  \cos \theta_{q\nu} 
=
  E_s / E_\nu |{\bf q}|
$
with 
$
  \cos \theta_{q \nu} 
=
  (m^2_s + |{\bf q}|^2 + E^2_\nu - E^2_s)
/
  2 E_\nu |{\bf q}|
$. The allowed range of the scattering angle, 
$\cos \theta_{q \nu} \leq 1$, gives the $|{\bf q}|$
integration range, $E_\nu - \sqrt{E^2_s - m^2_s} \leq
|{\bf q}| \leq E_\nu + \sqrt{E^2_s - m^2_s}$
with $E_s = E_{\nu} - \Delta E_{nl}$.
Putting things together, the differential cross
section of the recoil energy $T_r$ becomes,
\begin{eqnarray}
  \frac {d \sigma_{nl}}{d T_r}
& \equiv &
  \frac 1 {T_r}
  \int \frac{|{\bf q}| d |{\bf q}|}{4 \pi}
    \frac{1}{8 m_e^2 E_\nu^2 }
  \overline{|\mathcal M|^2}
  K_{nl}(T_r, {\bf q}).
\label{eq:sigma8}
\end{eqnarray}

The atomic effect can now be parameterized as the 
so-called $K$-factor,
\begin{equation}
    K_{nl}(T_r, {\bf q})
\equiv
  \frac 1 2 \frac{(2 m_e T_r)^{3/2}}{(2\pi)^3}
  \frac 1 {2 l + 1}
  \sum_{m,\,l'm'}
  \left| f^{T_r l'm'}_{nlm}({\bf q}) \right|^2.
\label{atomickernel}
\end{equation}
The $1/(2 l + 1)$ factor accounts for the average
over the initial magnetic quantum number $m$.
Usually, the direct detection experiments can only distinguish the 
energy deposit but not the angular quantum numbers $m$ for the
initial- and $\{l', m'\}$ for the final-state electrons.
So these quantum numbers are summed over when defining the
$K$-factor in \geqn{atomickernel}. For completeness, we rewrite
the $K$-factor more explicitly \cite{Essig:2012yx},
\begin{eqnarray}
  K_{nl}(T_r, |{\bf q}|)
=
  \frac{(2 m_e T_r)^{3/2}}{2 (2\pi)^3}
  \sum_{l'L} (2l'+1) (2L+1)
\left| \int r^2 dr j_L(|{\bf q}| r)  R^*_{T_r l'}(r) R_{nl}(r) \right|^2
\begin{pmatrix}
  l'  &  L  &  l \\
  0  &  0  &  0 
\end{pmatrix}^2,
\label{eq:kernel_ionization}
\end{eqnarray}
with the wave functions divided into radial and angular parts,
$\phi_{nlm}({\bf r}) = R_{nl} (r) Y_{lm} (\hat{r})$ and
$\phi_{T_r l'm'}({\bf r}) = R_{T_r l'} (r) Y_{l'm'} (\hat{r})$.
For convenience, the spatial integration variable
${\bf y}$ has been replaced by ${\bf r}$ and $r$ is radius.
The Bessel function $j_L(|{\bf q}| r)$ originates from the exponential
$e^{i {\bf q} \cdot {\bf y}}$ of \geqn{formfactor-1}.
After integrating over the solid angles, only the radial functions \cite{Bethe:1957ncq}
$R_{nl}$ for the bound and $R_{T_r l'}$ for the ionized electrons
survive. Most importantly, the $K$-factor becomes no
longer a function of ${\bf q}$ but its magnitude $|{\bf q}|$.

Note that the normalization of $K$-factor varies in literature
\cite{Essig:2011nj,Essig:2015cda,Catena:2019gfa,Chen:2021ifo}.
In our definition \geqn{eq:kernel_ionization},
the summation over $m$ and $m'$ gives a factor of $2 l + 1$
and $2 l' + 1$, respectively. That $2l+1$ factor has been
canceled with the one in \geqn{atomickernel}.
With this convention, \geqn{eq:kernel_ionization} 
defines the average $K$-factor for a single electron 
in the $nl$ state and \geqn{eq:sigma8} is
the cross section per electron. 
In principle, the final-state angular quantum number 
$l'$ should sum up to infinity. 
However, in practice the summation is cut off at 
sufficiently large $l' \sim \mathcal{O} (100)$ 
and higher contributions are neglected \cite{Catena:2019gfa, Chen:2021ifo}
\footnote{
\label{footnote1}
Useful python codes can be found on
\href{https://github.com/temken/DarkARC}{https://github.com/temken/DarkARC} and
\href{https://github.com/XueXiao-Physics/AtomIonCalc}{https://github.com/XueXiao-Physics/AtomIonCalc}.}. 

To see the basic features of the $K$-factor more
clearly, let us compare the scattering with free and
bound electrons. If the electron is not tightly bound in
the atom, or equivalently $n \rightarrow \infty$, the whole
process should reduce to the scattering with a free
electron. The differential cross section of the
scattering with a free electron is, 
\begin{eqnarray}
  \frac {d \sigma^0}{d T_r}
=
    \frac{1}{T_r}
    \int |{\bf q}|^2 d|{\bf q}|
    \frac{\overline{|{\mathcal M}|^2}}{64 \pi  m_e^2 |{\bf p}_\nu|^2}
    \delta(|{\bf p}| - |{\bf q}|)
=
  \frac {|\mathcal M|^2}
        {32 \pi m_e |{\bf p}_\nu|^2},
\label{eq:xsec-free}
\end{eqnarray}

\begin{figure}[t]
\centering
\includegraphics[width=8.2cm]{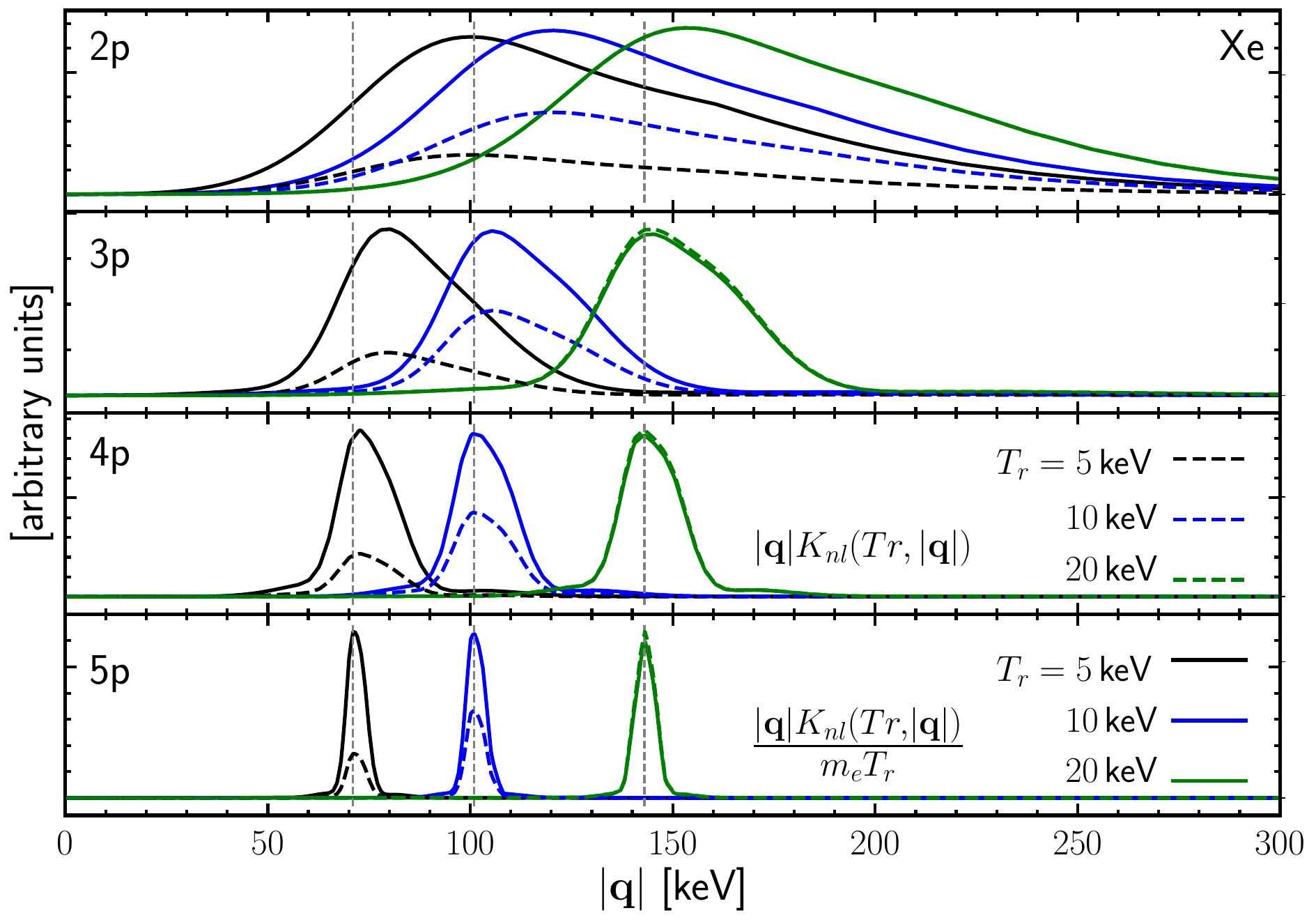}
\quad
\includegraphics[width=8.5cm]{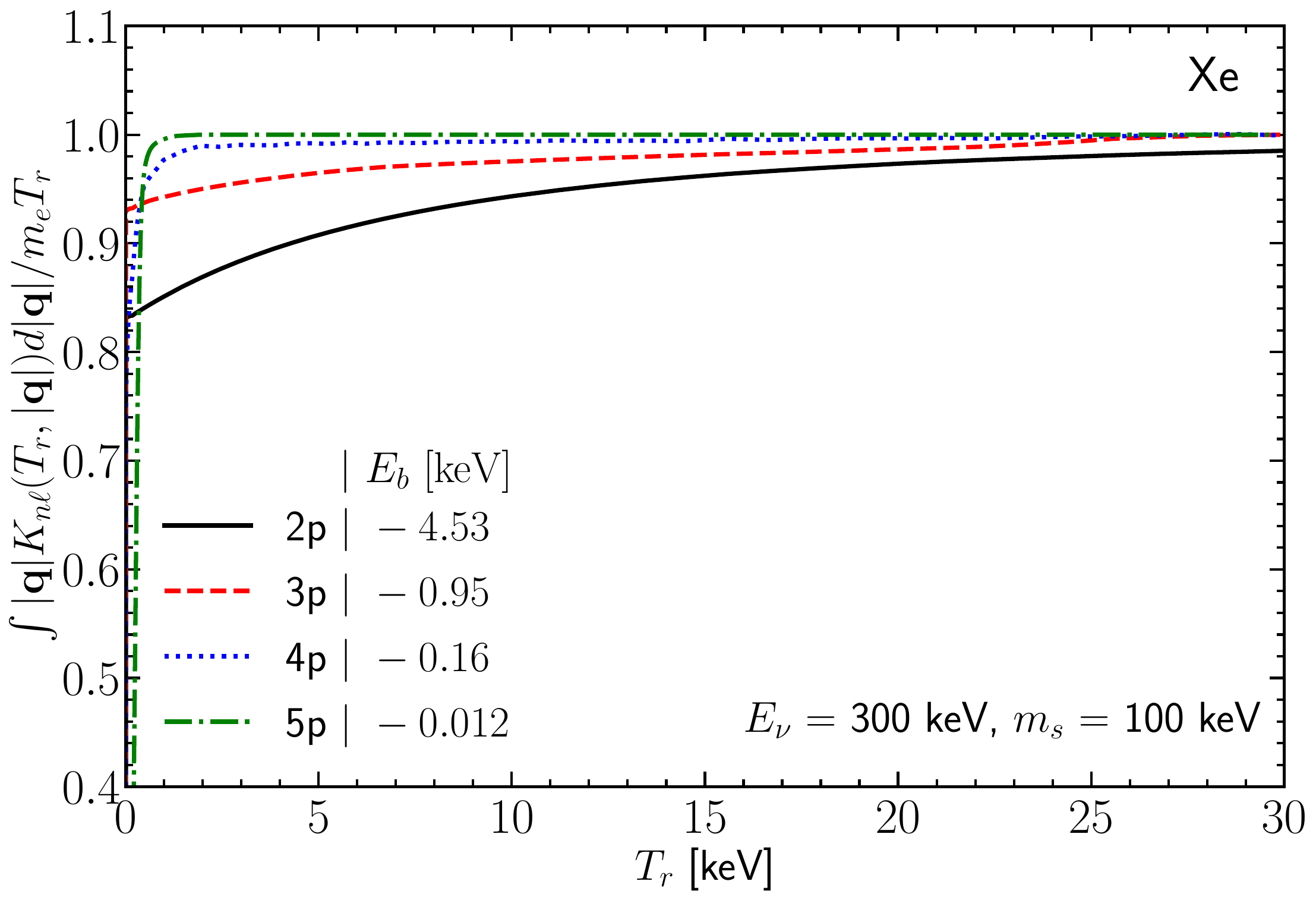}
\caption{
The left panel shows $|{\bf q}| K_{nl} (T_r, |{\bf
q}|)$ (dashed) and the phase space ratio $ |{\bf q}|
K_{nl}(T_r, |{\bf q}|) / m_e T_r$ 
(solid) for the $2p$, $3p$, $4p$ and $5p$ orbitals
from top to bottom. Three typical recoil energies
$T_r = 5\,$keV (black), 10\,keV (blue), and $20\,$keV
(green) are shown for illustration. The peak locations,
$|{\bf q}| = \sqrt{2 m_e T_r}$, estimated from the
scattering with a free
electron are shown as vertical gray lines.
As a function of the electron recoil energy $T_r$, 
the right panel shows the phase space ratio $\int |{\bf q}|
K_{nl}(T_r, |{\bf q}|) d |{\bf q}| / m_e T_r$
for the $2p$ (black solid), $3p$ (red dashed), 
$4p$ (blue dotted) and $5p$ (green dash-dotted)
orbitals in the Xe atom with solar neutrino energy $E_\nu = 300$\,keV
and the sterile neutrino mass $m_s = 100$\,keV.
}
\label{fig:PhaseSpaceRatio}
\end{figure}

with $|{\bf p}|^2 = 2 m_e T_r$. For easy comparison, we
have kept the momentum transfer integration which can be
removed by the $\delta(|{\bf p}| - |{\bf q}|)$ function.
Comparing \geqn{eq:xsec-free} with \geqn{eq:sigma8}, we
can see that the $K$-factor reduces to
$|{\bf q}| \delta(|{\bf p}| - |{\bf q}|)/2$
in the free electron scattering limit.
For the scattering with a free electron, the
two-body phase space reduces to a single integration
over either the recoil energy $T_r$ or equivalently the
momentum transfer $|{\bf q}|$.
But for the bound electron case, the phase space has
double integration, over $T_r$ and $|{\bf q}|$ which are
no longer correlated. This is because the initial bound
electron does not have definite momentum but a 
distribution. Consequently, the contribution from different
momentum transfer should be integrated.
Besides $\overline{|\mathcal M|^2} / 32 \pi m^2_e |{\bf
p}_\nu|^2 T_r$, the phase space integrations are $m_e
T_r$ and $\int |{\bf q}| K_{nl}(T_r, |{\bf q}|) d |{\bf
q}|$ for the free and bound electron cases,
respectively. Assuming constant $|\mathcal M|^2$, the
atomic enhancement can be roughly measured by the ratio
between the phase space integrations, $\int |{\bf q}|
K_{nl}(T_r, |{\bf q}|) d |{\bf q}| / m_e T_r$.

The left panel of \gfig{fig:PhaseSpaceRatio} shows the atomic
factor $|{\bf q}| K_{nl}(T_r, |{\bf q}|)$ (dashed) and the
atomic factor ratio $|{\bf q}| K_{nl}(T_r, |{\bf q}|) /
m_e T_r$ (solid) as functions of the momentum transfer
$|{\bf q}|$. The curves are narrower for outer shell
electrons (larger $n$) and the typical width keeps
growing when $n$ becomes smaller. 
The half width at the half height
$\sigma_{|{\bf q}|} = (50, 20, 10, 4)$\, keV
for the ($2p$, $3p$, $4p$, $5p$) electron
can be 
directly read off from the solid lines.
The solid and dashed lines share the same width
since the only difference between them is a constant
factor. For given electron shell, the width
is independent of $T_r$.
The width is exactly a manifestation of the electron motion
inside atom.
Pointing in all directions, the initial electron
momentum smears the momentum transfer
$|{\bf q}|$ and hence the recoil energy $T_r$. 
The larger initial momentum, the larger smearing effect. 
With binding energy
$|E_b| = (4.53, 0.95, 0.16, 0.012)$\,keV for 
($2p$, $3p$, $4p$, $5p$) orbitals \cite{Catena:2019gfa}, 
the initial electron momentum can be estimated 
by non-relativistic dispersion relation,
$|{\bf p}_i| \approx \sqrt{2 m_e|E_b|} = $(68, 31, 13,
3.5)\,keV correspondingly. These numbers roughly match 
the width read off from the curves. 
Although the bound electron motion broadens the curves, 
the central values are largely unaffected, 
especially for the outer shells. 
The peak position can be estimated by the momentum transfer 
$|{\bf q}^{\rm peak}|
\approx \sqrt{2 m_e T_r}$ of the free electron scattering.
For the three curves of $T_r = (5, 10, 20)$\,keV,
the estimated peak positions are $|{\bf q}^{\rm peak}|
\approx (71, 101, 143)$\,keV as indicated by three
vertical grey lines from left to right. With larger $n$,
the peak position becomes closer to the free electron 
momentum transfer.

The right panel of \gfig{fig:PhaseSpaceRatio} shows the
integrated phase space ratio as functions of the recoil energy $T_r$.
The ratio starts from a relatively small value for vanishing $T_r$
and converges to $1$ with increasing $T_r$. 
This happens because for small $T_r$ the allowed momentum
transfer is suppressed to have smaller phase space \cite{Essig:2011nj}.
In contrast, the electron approaches a free particle to have
larger phase space with increasing $T_r$. For larger
$n$, the integrated phase space ratio approaches $1$ faster,
since the electron is less tightly confined to the atom.

\begin{figure}[t]
\centering
\includegraphics[height=6.65cm,width=8.2cm]{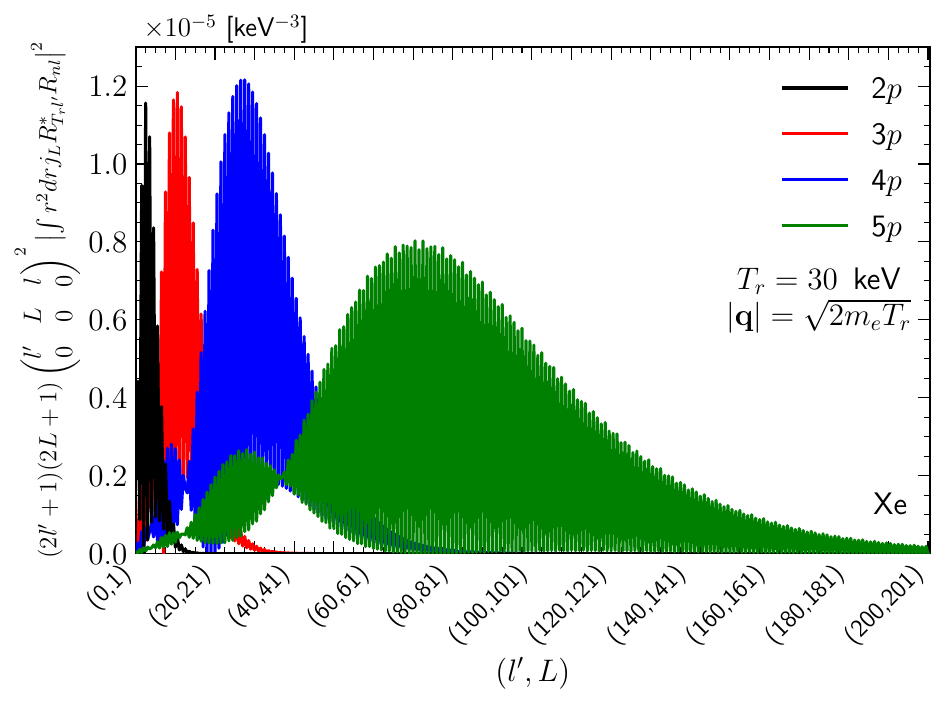}
\includegraphics[width=8.5cm]{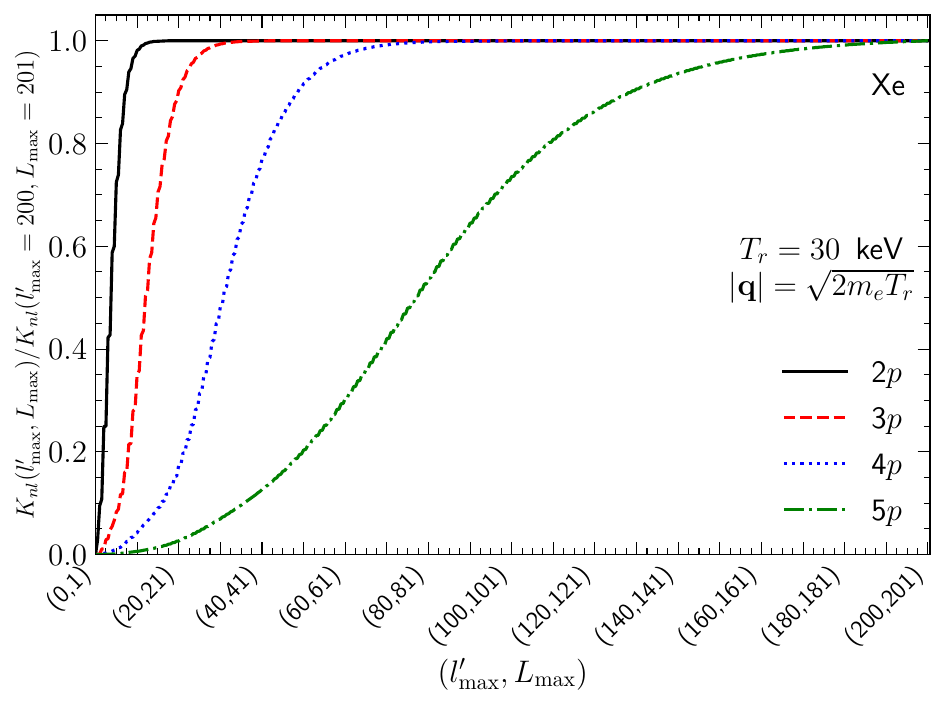}
\caption{
The left panel shows the non-zero terms 
of the double summation in (\ref{eq:kernel_ionization})
as a function of $l',L$. The right panel shows the fraction between the partial sum for a given set of $l'_{\rm max}, L_{\rm max}$ and the total sum with $l'_{\rm max} = 200, L_{\rm max} = 201$. In both cases, $T_r = 30$\,keV 
and $|{\bf q}| = \sqrt{2m_e T_r}$. The lines represents the 2p (black), 3p (red), 4p (blue) and 5p (green) orbitals in the Xe atom.
}
\label{fig:PartialSum}
\end{figure}

The left panel of \gfig{fig:PartialSum} 
shows the non-zero terms inside the double summation
of \geqn{eq:kernel_ionization}
as a function of the angular indices $(l',L)$ for the $2p$, 
$3p$, $4p$, and $5p$ orbitals. Those terms violating
$|l' \pm L| = l$ are zero and omitted in the plot for simplicity.
With increasing $n$, the peak shifts to 
the right. The $2p$ curve peaks around 
$l' = 5$ and drops to zero at $l' = 15$. For comparison,  
the $5p$ curve peaks around $l' = 95$ and is
non-zero even for $l' > 190$.
The exact evaluation of the $K_{nl}$ factor in 
\geqn{eq:kernel_ionization} requires  
double summation over the principal quantum numbers, 
$l'$ and $L$, to infinity which is highly time-consuming. 
To increase efficiency, we cut the summation at 
large enough $l'_{\rm max}$ and $L_{\rm max}$ to
guarantee precision.

The relative precision can be 
estimated from the right panel of \gfig{fig:PartialSum}. 
Each curve shows the ratio
between the partial sum up to ($l'_{\rm max}$, $L_{\rm max}$) divided by the one up to ($200$, $201$) 
for the $2p$, $3p$, $4p$, and $5p$ orbitals. 
With increasing ($l'_{\rm max}$, $L_{\rm max}$), 
the ratio converges to 1. As $n$ increases, 
larger $l'_{\rm max}$ and $L_{\rm max}$ are needed.
For the $2p$, $3p$, $4p$, and $5p$ orbitals  
to achieve sub-percentage precision,
($l'_{\rm max}$, $L_{\rm max}$) needs to be 
at least (10, 11), (30, 31), (70, 71), and (190, 191), respectively. 

\begin{figure}[t]
\centering
\includegraphics[scale = 0.4]{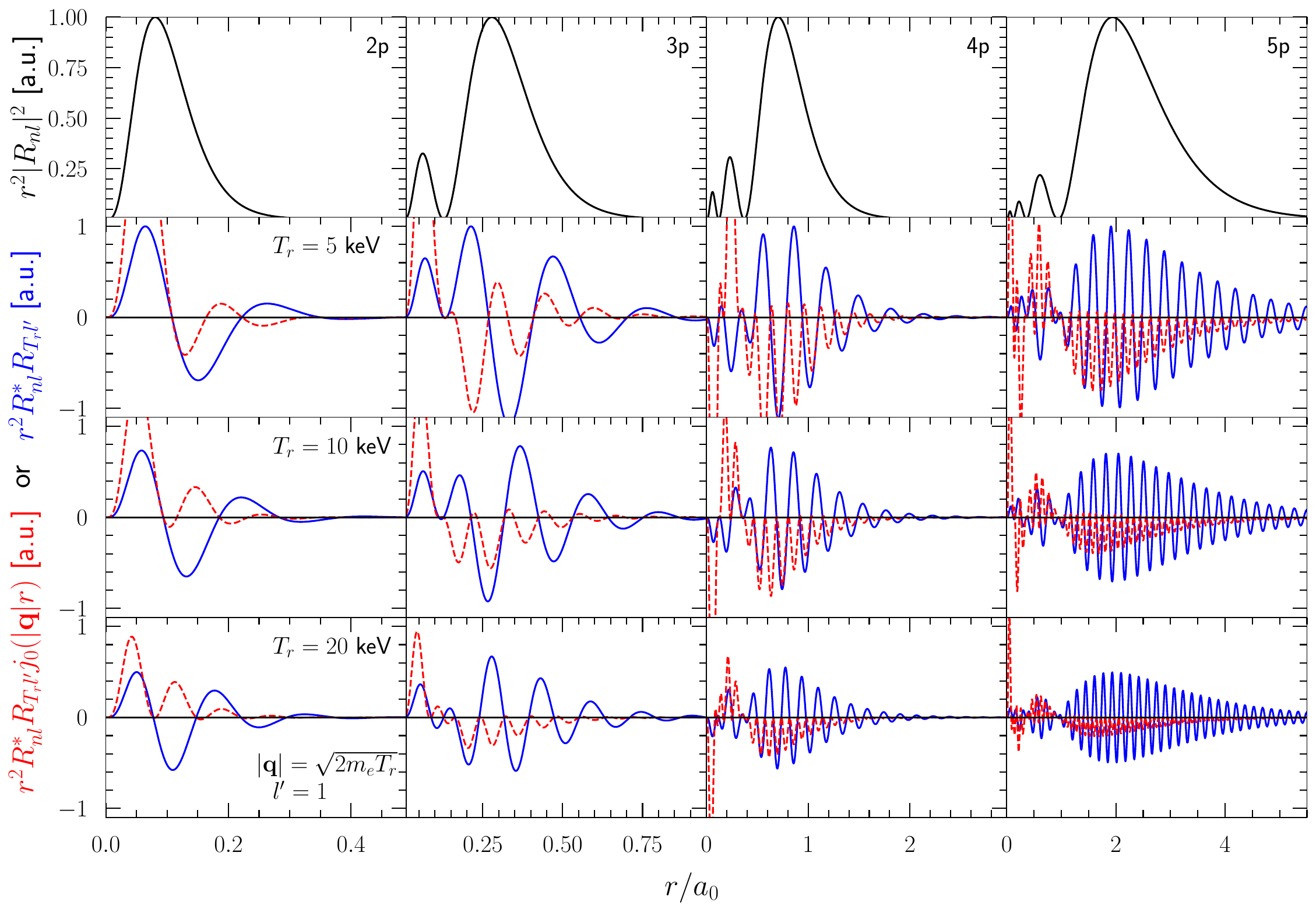}
\caption{The oscillatory behavior in the bound and ionized
electron wave functions. Each column stands for an individual
initial electron level, $(nl) = 2p$, $3p$, $4p$, and $5p$.
The first row is purely for the initial electron radial wave
function $r^2 |R_{nl}|^2$ which appears in the integration for
normalization. The next three rows are for the convolution
between the initial and final electron radial functions
$r^2 R^*_{nl} R_{T_r l'}$ (blue solid) and
$r^2 R^*_{nl} R_{T_r l'} j_L(|{\bf q}| r)$ with
$L = 0$ (red dashed) for
electron recoil energies $T_r = (5, 10, 20)$\,keV.
For illustration, we take the peak value $|{\bf q}| = \sqrt{2 m_e T_r}$
of the momentum transfer.
The horizontal axises are in the unit of Bohr radius $a_0 = 1/m_e \alpha$.}
\label{fig:grid_wave_function}
\end{figure}
The integrand in \geqn{eq:kernel_ionization} 
is highly oscillatory and 
illustrated in \gfig{fig:grid_wave_function}. The top panels
show purely the initial bound electron wave functions as a
convenient combination $r^2 |R_{nl}|^2$ (black solid) that
should appear in normalization integration. While the initial
wave functions are quite regular with only one major peak, the
final ionized electron wave functions oscillate a lot. This
oscillatory behavior clearly appears in the product
combination $r^2 R^*_{nl} R_{T_r l'}$ (blue solid). Since the
bound and ionized states are orthogonal to each other, the
product oscillates symmetrically around the vanishing value so
that the integration should also vanish. The third
component, the Bessel function $j_L(|{\bf q}| r)$, enters
the atomic form factor to render a nonzero $K$-factor
\geqn{eq:kernel_ionization}. The full combination
$r^2 R^*_{nl} R_{T_r l'} j_L(|{\bf q}| r)$ (red dashed) has
asymmetric oscillation so that its integration can be
nonzero.
With larger $n$, $l$, and $T_r$,
the oscillation becomes more frequent.
To accurately perform
the integration in \geqn{eq:kernel_ionization}, a large number
of grid points over $r$ is necessary.
A sub-percentage precision requires $3000$ 
grid points in the range $r \in [0, 10 a_0]$ where 
$a_0 \equiv 1/m_e \alpha$ is the Bohr radius.

\section{Neutrino Scattering into A Massive Sterile Neutrino with Light Mediators}
\label{sec:CrossSections}

With the second quantization formalism for the atomic effect
established in the previous section,
we can proceed to calculate the recoil spectrum \geqn{eq:sigma8}
by implementing the concrete scattering matrix element
$\overline{|\mathcal M|^2}$ as input from the particle physics side.
For the solar neutrino scattering into a massive sterile neutrino
in the final state, we first derive the differential cross
section in \gsec{sec:xsec} and then show the modification
due to atomic effects in \gsec{sec:enhancement}.

\subsection{Neutrino Scattering with Free or Bound Electrons}
\label{sec:xsec}

The XENON1T excess in the low energy region can be
explained by the solar neutrino scattering with electron
via a scalar or pseudo-scalar mediator
\cite{Ge:2020jfn}.
\begin{equation}\label{eq:scalar_lagrangian}
    \mathcal{L}_{\mathrm{int}}
\equiv
    \bar{\nu}
    \left(y_{S}^{\nu}
    +\gamma_{5} y_{P}^{\nu}\right) 
    \phi \nu_{s}
    +
    \bar{e}
    \left(y_{S}^{e}+\gamma_{5} y_{P}^{e}\right) 
    e \phi+h . c .
\end{equation}
In addition to the recoil electron, a massive sterile
neutrino appears in the final state. For the scattering
with a free electron, the cross section is a function of
the recoil energy $T_r$,
\begin{subequations}
\begin{eqnarray}
  \mbox{Scalar}
& : &
  \frac{d \sigma^0_S}{d T_r} 
=
  \frac {(y^\nu_S y^e_S)^2 (2m_e + T_r)}
  {8 \pi E_\nu^2 }
  \frac {2 m_e T_r + m^2_s}
  		{(2 m_e T_r + m^2_\phi)^2},
\label{eq:XsecFreeScalar}
\\
  \mbox{Pseudo-Scalar}
& : &
  \frac{d \sigma^0_P}{d T_r} 
=
\frac {(y^\nu_P y^e_P)^2 T_r}
  {8 \pi E_\nu^2 }
  \frac {2 m_e T_r + m^2_s}
  		{(2 m_e T_r + m^2_\phi)^2},
\end{eqnarray}
\label{dcsfree}
\end{subequations}
In addition to
the neutrino energy $E_\nu$, $m_s$ and $m_\phi$ are the
sterile neutrino and scalar/pseudo-scalar mediator
masses, respectively. Contrary to the common
expectations \cite{Boehm:2020ltd} with massless
neutrinos, a massive sterile and a light mediator
($m^2_\phi \ll 2 m_e T_r \ll m^2_s$) can introduce $1 /
T_r$ enhancement even for the pseudo-scalar mediator
\cite{Ge:2020jfn} by a factor of $m^2_s / 2 m_eT_r$. 
For a low energy electron recoil signal, 
such as the XENON1T excess at 
$T_r \approx (2 \sim 3)\,\mbox{keV}$, 
the mediator mass should satisfy an upper 
bound $m_\phi \ll (45 \sim 55)\,\mbox{keV}$ 
to receive enhancement.

For the scattering with a bound electron, the electron
part resembles the free case for the scalar-type vertex
and receives a momentum insertion for the pseudo-scalar
one as shown in \geqn{eq:M},
\begin{subequations}
\begin{eqnarray}
    \overline{|{\mathcal M}_S|^2}
& = &
    \frac{(y^\nu_S y^e_S)^2}{2 (q^2 - m_\phi^2)^2}
    \tr[\slashed p_\nu (\slashed p_{\nu_s} + m_s)]
    \tr[(\fp_e + m_e) (\fp'_e + m_e)],
\\
    \overline{|{\mathcal M}_P|^2}
& = &
    \frac{(y^\nu_P y^e_P)^2}{2 (q^2 - m_\phi^2)^2}
  \frac 1 {4 m_e^2}
  \tr [\slashed p_\nu \gamma_5 (\slashed p_{\nu_s} + m_s) \gamma_5]
  \tr [ (\fp_e + m_e)
    ({\bf q} \cdot \boldsymbol \gamma) \gamma_5
    (\fp'_e + m_e)
    ({\bf q} \cdot \boldsymbol \gamma) \gamma_5 ].
\label{Mforscalar}
\end{eqnarray}
\end{subequations}
When calculating the trace, we should remember that 
the spinor for electron $u(m_e)$ is the one without
momentum. But it does not mean this approximation has no
information of the electron momentum. Actually, the
electron momentum appears as the gradient operator in
\geqn{DiracSolution1}, which when acting on the
integration will generate the momentum transfer
${\bf{q}}$. This is also the origin of $\textbf{q} \cdot
\boldsymbol \gamma$ in the pseudo-scalar matrix element.
Putting the scalar and pseudo-scalar cases together, the
spin-averaged matrix element is,
\begin{equation}
  \overline{|\mathcal M_{S,P}|^2}
=
    2 \left(
    4 m_e^2 |y^\nu_S y^e_S|^2
+ |{\bf q}|^2 |y^\nu_P y^e_P|^2 
\right)
  \frac {|{\bf q}|^2 + m^2_s - \Delta E^2_{nl}}
  			{(|{\bf q}|^2 + m^2_\phi - \Delta E^2_{nl})^2}.
\label{eq:M2-solarNu}
\end{equation}
Although derived for the bound electron case,
\geqn{eq:M2-solarNu} can reproduce the leading terms of
the free electron case
\geqn{dcsfree} with the following replacements.
The initial electron
is at rest and the energy difference is exactly the recoil kinetic
energy, $\Delta E_{nl} = T_r$, of the final electron. In addition,
the momentum
transfer magnitude $|{\bf q}|$ is uniquely related to the recoil
energy, $|{\bf q}|^2 = T^2_r + 2 m_e T_r$, due to the on-shell
condition of the final-state electron.

On the other hand, a bound electron in the initial state has
no definite momentum. There is no way to use the on-shell
condition for the final-state ionized electron to correlate
the energy change $\Delta E_{nl} \equiv T_r - E_{nl}$ and the
momentum transfer magnitude
$|{\bf q}|$. The differential cross section \geqn{eq:sigma8}
is then an integration over all the possible momentum transfers,
\begin{subequations}
\begin{eqnarray}
  {\rm Scalar}
& : & 
  \frac {d \sigma^{nl}_S}{d T_r}
=
  \frac {(y^\nu_S y^e_S)^2} { 4 \pi E_\nu^2 T_r}
  \int |{\bf q}| d |{\bf q}|
  \frac {|{\bf q}|^2 + m^2_s - \Delta E^2_{nl}}
  			{(|{\bf q}|^2 + m^2_\phi - \Delta E^2_{nl})^2}
  K_{nl}(T_r, |{\bf q}|),
\\
  \mbox{Pseudo-Scalar}
& : &
  \frac {d \sigma^{nl}_P}{d T_r}
=
 \frac {(y^\nu_P y^e_P)^2} {4 \pi E_\nu^2 T_r}
  \int |{\bf q}| d |{\bf q}|
    \frac{|{\bf q}|^2}{4m_e^2}
  \frac {|{\bf q}|^2 + m^2_s - \Delta E^2_{nl}}
  			{(|{\bf q}|^2 + m^2_\phi - \Delta E^2_{nl})^2}
  K_{nl}(T_r, |{\bf q}|).
\end{eqnarray}
\label{dxsecBound-scalar}
\end{subequations}

The similar scenario
with a light vector boson mediator can also explain the XENON1T
excess \cite{Ge:2020jfn}.
For simplicity, we consider only the situation where 
$Z'$ couples to the left-handed neutrino,
\begin{equation}
    \mathcal{L}_{\text {int }}
\equiv
    g_{L}^{\nu} \bar{\nu} \gamma^{\mu} P_{L} \nu_{s}
    Z_{\mu}^{\prime}
    +
    \bar{e}\left(g_{V}^{e}-g_{A}^{e} \gamma_{5}\right) \gamma^{\mu} e Z_{\mu}^{\prime},
\end{equation}
while the electron coupling can have either vector ($g^e_V$)
or axial-vector ($g^e_A$) current coupling with $Z'$.
The differential cross section of neutrino scattering with
a free electron is,
\begin{equation}
    \frac{d \sigma^0_{V,A}}{d T_{r}}
=
  \frac {\left(g_{L}^{\nu} g_{V, A}^{e}\right)^{2}}
        {4 \pi E_{\nu}^{2}}
  \frac {4 m_e E^2_\nu - (2 m_e T_r + m^2_s) (2 E_\nu \pm m_e - T_r)}
        {\left(2 m_e T_r + m_{Z'}^2 \right)^2},
\label{XsecFreeVector}
\end{equation}
where the $+\,(-)$ signs are for the vector and axial-vector 
interactions, respectively. The sterile neutrino mass term
$m^2_s$ and hence the second term in the numerator
can be important only when $m_s$ becomes comparable with the
electron mass $m_e$ and the neutrino energy $E_\nu$. In this
paper, we focus on the mass region $m_s \sim \mathcal O(100)$\,keV.
Then with $4 m_e E^2_\nu$ dominating
the numerator, a $1/T^2_r$ enhancement is possible for a
light enough mediator, $m^2_{Z'} \ll 2 m_e T_r$.
For comparison, the differential cross section of neutrino
scattering with a bound electron is,
\begin{eqnarray}
  \frac {d \sigma_{V,A}^{nl}}{d T_r}
=
  \frac {(g^\nu_L g^e_{V,A})^2} {4 \pi E_\nu^2 T_r}
  \int |\boldsymbol{q}| d|\boldsymbol{q}|
  \frac {  4 E_\nu (E_\nu - \Delta E_{nl})
        \mp
          (|\boldsymbol{q}|^2 + m_s^2 -\Delta E^2_{nl})}
        {(|\boldsymbol{q}|^2 + m_{Z'}^2 - \Delta E^2_{nl})^2}
  K_{nl}(T_r, |{\bf q}|),
\label{dxsecBound-vector}
\end{eqnarray}
with $\mp$ for the vector and axial-vector couplings,
respectively.

\subsection{The Cross Section Enhancement from Atomic Effects}
\label{sec:enhancement}

\begin{figure}[t]
\centering
\includegraphics[width=8cm]{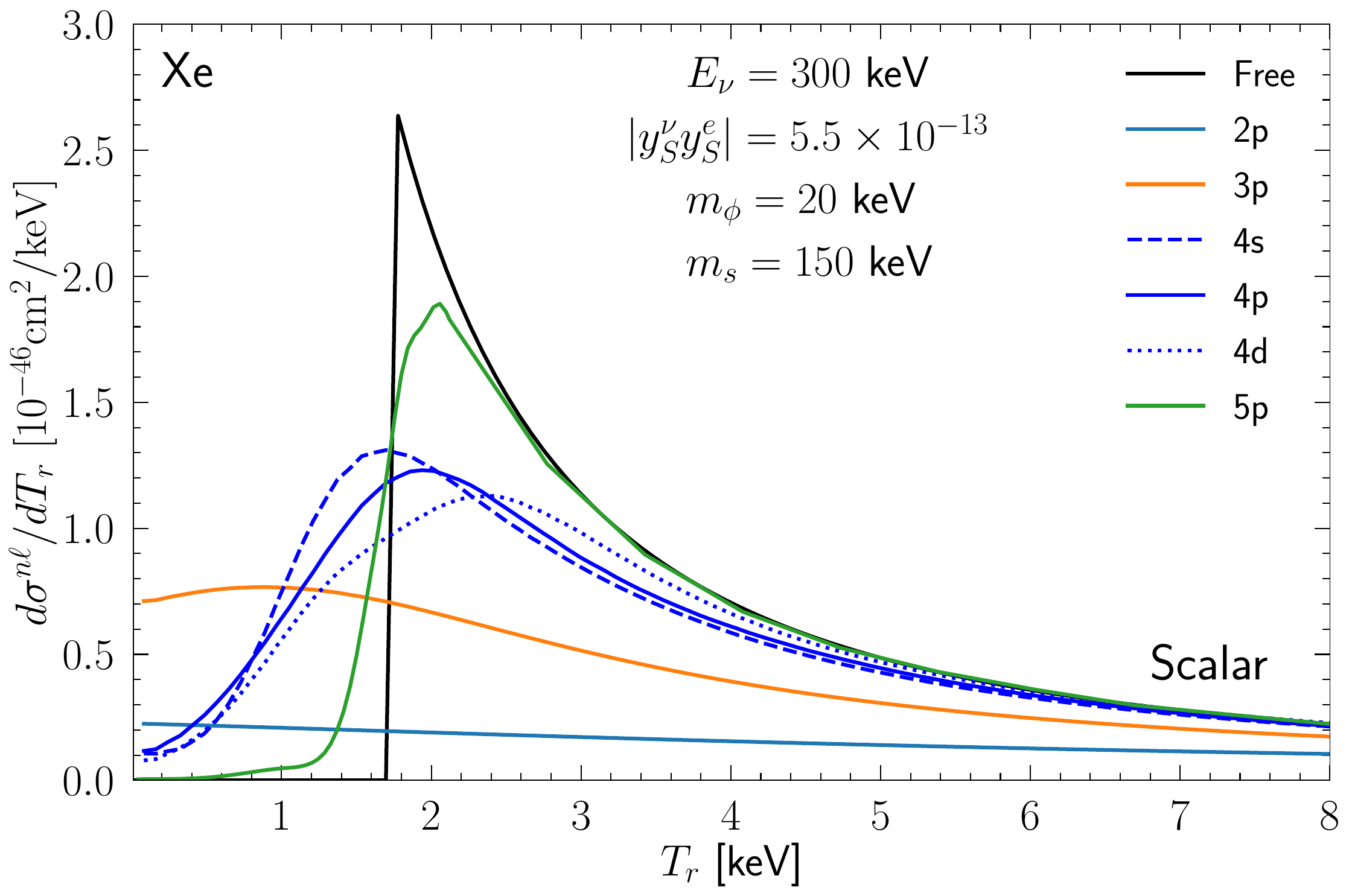}
\qquad
\includegraphics[width=8cm]{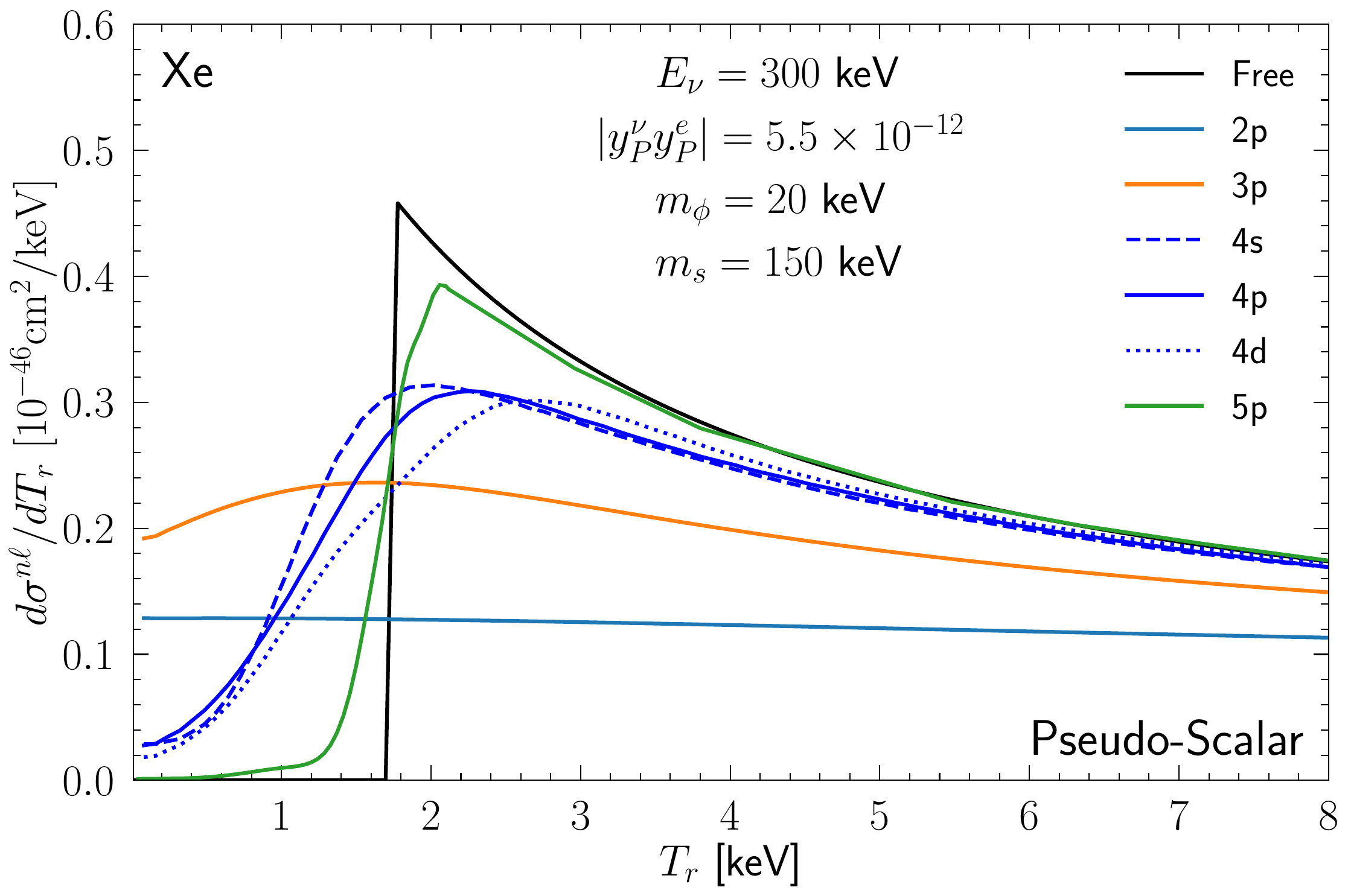}
\caption{The differential cross section of $\nu$-$e$ scattering 
with a massive sterile neutrino in the final state and
light scalar (left) or pseudo-scalar (right) mediator.
The black line is for the scattering with a free electron while
the colorful lines are for the bound electrons
$2p$ (solid cyan), $3p$ (solid yellow), $4s$ (dashed blue),
$4p$ (solid blue), $4d$ (dotted blue), and $5p$ (solid green)
of a Xenon atom. In both subplots, the final-state sterile neutrino
mass is $m_s = 150$ keV and the mediator mass is $m_\phi = 20$ keV.
}
\label{fig:Scalar_all_shells}
\end{figure}

By implementing the $K$-factor elaborated in
\gsec{sec:atomic}, we obtain the differential cross
sections \geqn{dxsecBound-scalar} and
\geqn{dxsecBound-vector} for scalar and vector
mediators. To see the features clearly, we first
evaluate $d \sigma^{nl}/ d T_r$ for the transition of a single
initial electron in a given $nl$ bound state to the
ionized electron with recoil energy $T_r$, shown as colorful lines in
\gfig{fig:Scalar_all_shells}. For comparison, we also
show the free electron case as black solid line.
We take the scalar and pseudo-scalar mediators
to illustrate the physics picture while the vector
case is similar.

The difference is quite significant. Especially, the
smearing effect makes the sharp peak much fatter and lower
mainly due to the electron motion in the atom.
For smaller principal quantum number $n$, the
reduction is much stronger. One reason is the phase space
ratio due to the atomic $K$-factor as shown in \gfig{fig:PhaseSpaceRatio}. 
Another factor comes from the scattering matrix element.
The electron in the atomic Coulomb potential 
has a negative binding energy $E_b $ which is exactly $ E_{nl}$. 
Intuitively, It is more difficult for an electron 
with larger binding energy to be recoiled off. 
If the energy change $\Delta E_{nl} = T_r - E_{nl}$ in \geqn{eq:M2-solarNu} dominates, 
especially for the inner shells, 
the scattering matrix element can be greatly suppressed.
This is because the momentum transfer peaks at
$|{\bf q}| = \sqrt{2 m_e T_r}$ with large spread above the
peak position as shown in \gfig{fig:PhaseSpaceRatio}.
So the modification of the differential cross section is a combined
result of the $K$-factor phase space integration reduction
and the scattering matrix element suppression. 
On the other hand, 
the electron in outer shells is loosely trapped with 
marginal suppression and is very similar to the free case.

In addition, the smearing effect extends the spectrum beyond the cut-off.
For free electron scattering, the recoil  energy is bounded from below,
$T_r \geq T^-_r$ as defined in our earlier paper
\cite{Ge:2020jfn}. A nonzero lower limit $T^-_r \approx
m^4_s / 8 m_e E^2_\nu$ arises when expanding the
sterile neutrino mass to the fourth power. For the
$E_\nu = 300\,$keV used to plot the black solid
line, the differential cross section vanishes around
$1.7$ keV. However, bound electrons have distributed
momentum in all directions and hence can smear the
momentum transfer to extend the 
recoil energy $T_r$ even down to $0\,$keV. 
Furthermore, the $2p$ or $3p$ electrons even 
have a non-zero differential cross section at vanishing $T_r$.
As shown in \geqn{eq:sigma8}, the phase space integration is
$|{\bf q}| d |{\bf q}|$. Since the lower limit
$E_\nu - \sqrt{(E_\nu - \Delta E_{nl})^2 - m^2_s}$ of $|{\bf q}|$
with $\Delta E_{nl} > 0$ is always nonzero for the scattering of bound
electron into an ionized one, the phase space would not disappear even for
$T_r = 0$\,keV. This behavior can also explain the
sudden drop of the $5p$ curve when approaching vanishing $T_r$
in \gfig{fig:PhaseSpaceRatio}.

\begin{figure}[t]
\centering
\includegraphics[width=8.4cm]{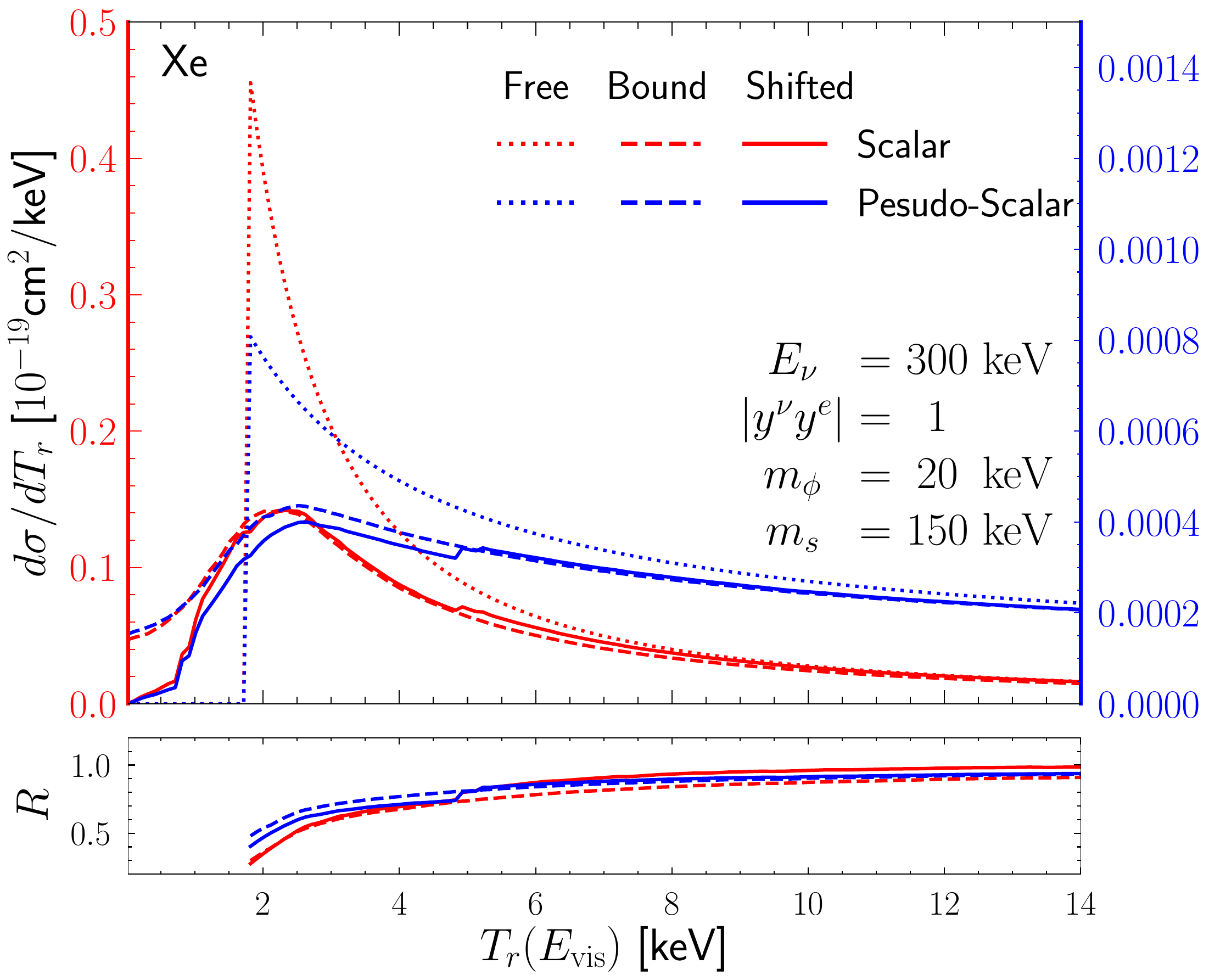}
\qquad
\includegraphics[width=8.4cm]{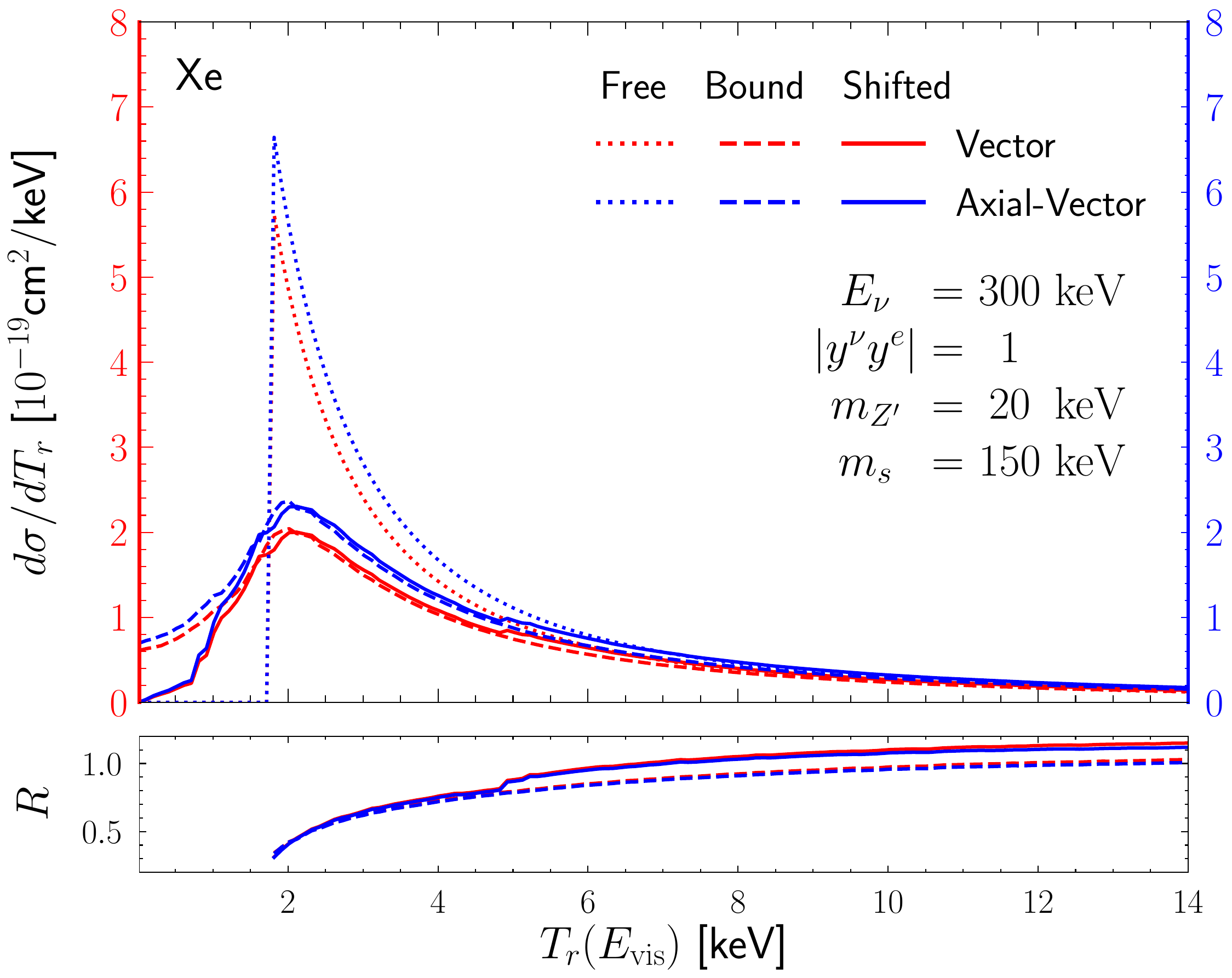}
\caption{
The differential cross section of neutrino
scattering with all electrons of Xe atom into a massive
sterile neutrino. The left panel shows the case with
scalar (red) and pseudo-scalar (blue)
mediators. For comparison, we also show the vector mediator
case in the right panel with vector (red)
and axial vector (blue) couplings.
Both panels show the results with free (dotted), bound (dashed),
and the shifted one including recombination energies (solid).
The enhancement from the free electron scattering cross section
is shown as ratio $R$ in the lower panels for comparison.
Since the couplings only appear as overall factor, we adopt
$|y^\nu y^e| = |g^\nu g^e| = 1$ for simplicity.
}
\label{fig:total_diff_ratio}
\end{figure}

Solar neutrino can scatter with electrons of
different quantum numbers inside the Xenon atom.
For the free electron scattering,
the total $d \sigma / d T_r$ is simply $N_e = 54$ times of
the differential cross section for a single electron where
$N_e$ is the electron number in the Xenon atom. The total cross
section for the bound electron scattering is a sum over all
the differential cross sections $d \sigma^{nl} / d T_r$,
\begin{eqnarray}
  \mbox{Free}
~ : ~
  \frac{d \sigma}{d T_r}
=
  54 \times \frac{d \sigma^0}{d T_r},
\quad {\rm vs} \quad 
  \mbox{Bound}
~:~
  \frac{d\sigma_{A}}{dT_r}
=
  \sum_{nl} 2(2 l + 1) \frac{d \sigma^{nl}}{dT_r}.
\label{totalenergyspectrum}
\end{eqnarray}
Of the weight $2(2 l + 1)$,
the factor 2 for the bound case accounts 
for the two electron spin degrees of freedom 
with the same $nlm$ quantum numbers. 
Note that the summation over the magnetic quantum 
number $m$ results in the $2 l + 1$ factor. 
As illustrated by the blue lines in \gfig{fig:Scalar_all_shells},
the $4s$, $4p$, and $4d$ curves with the same  principal
quantum number $n = 4$ but different angular quantum numbers
have similar shapes and amplitudes.
More generally, orbitals with the same principal quantum 
number have roughly the same differential cross sections.
The $n = 5$ orbitals have higher and 
sharper peaks than the $n = 4$ ones. However, the 
electron number of the former, $2 (5s) + 6 (5p) = 8$, 
is less than half of the later, $2 (4s)
+ 6 (4p) + 10 (4d) = 18$. So the total
differential cross section of atomic scattering 
(thick dashed and solid)
in \gfig{fig:total_diff_ratio} is 
mainly contributed by the $4s$, $4p$, and $4d$ curves (blue) 
in \gfig{fig:Scalar_all_shells}. 
Since the differential cross section $d \sigma^{nl} /
d T_r$ for a single electron is reduced for all $nl$ 
quantum numbers, the total result $d \sigma_A / d 
T_r$ is also reduced from the free electron case. 
As shown in the lower panel, the reduction 
is roughly a factor of 0.5  
in the low-energy region ($T_r \lesssim 5$ keV) and 
gradually recovers to 1 in 
the high energy region ($T_r \gtrsim 10$ keV) 
for the scalar or pseudo-scalar mediators. 
For vector and axial-vector interactions, atomic effects bring the same features.

A key difference between the scalar/pseudo-scalar
and vector/axial-vector interactions is that the latter has
much larger cross section by almost a factor of $10 \sim 15$,
as shown in the upper panels of \gfig{fig:total_diff_ratio}. 
As discussed below
\geqn{XsecFreeVector}, the matrix element part is approximately
$4 m_e E_\nu^2 / (2 m_e T_r + m^2_{Z^\prime})^2$
for the free electron scattering with vector mediator. 
For comparison, the scalar case has 
$m_e (2 m_e T_r + m^2_s) / (2 m_e T_r + m^2_\phi)^2$.
With tiny mediator mass,
$m^2_\phi, m^2_{Z'} \ll 2 m_e T_r$, the difference mainly
comes from the numerator part. The ratio between 
vector and scalar matrix elements is
$4 m_e E_\nu^2 / m_e (2 m_e T_r + m^2_s)$.
We can see that the vector
case has a major contribution $4 m_e E^2_\nu$.
With $E_\nu \approx 300$\,keV, $m_s \approx 150$\,keV,
and $|{\bf q}|^2 \sim 2 m_e T_r \approx 3000$\,keV,
the vector cross section is naturally enhanced by a factor
around 10.

In addition to the free (dotted) and bound (dashed)
electron curves, \gfig{fig:total_diff_ratio} also shows
the shifted results. This is because the energy deposit
$\Delta E_{nl}$ is not just the electron recoil energy $T_r$
but also the binding energy $E_b = E_{nl}$. The later is
released when an ambient electron is attracted by the
positively charged Xenon atom to fill the hole left by
the ionized electron \cite{Szydagis:2011tk}.
The differential cross section is then shifted to the
right as the solid lines. The binding energy is
typically $\mathcal{O}(0.01)$ keV for outer shells
\cite{Szydagis:2011tk} and can be ignored in comparison
to the recoil energy. But the binding energy can be
as large as $\mathcal{O} (10)$\,keV for most inner
shells and hence seems able to induce significant
horizontal shift, 
the differential cross section $d \sigma^{nl} / d
T_r$ is suppressed by the same large binding energy in
the first place as discussed above. So the energy shift
does not affect the total differential cross section
$d \sigma / d E_{\rm vis}$ much as shown in the plot.

%\section{Constrains from Dark Matter Experiments}
%\label{Section:Constrains}

\section{Experimental Constraints}
\label{sec:Experiments}

\begin{figure}[t]
\centering
\includegraphics[width=8.4cm]{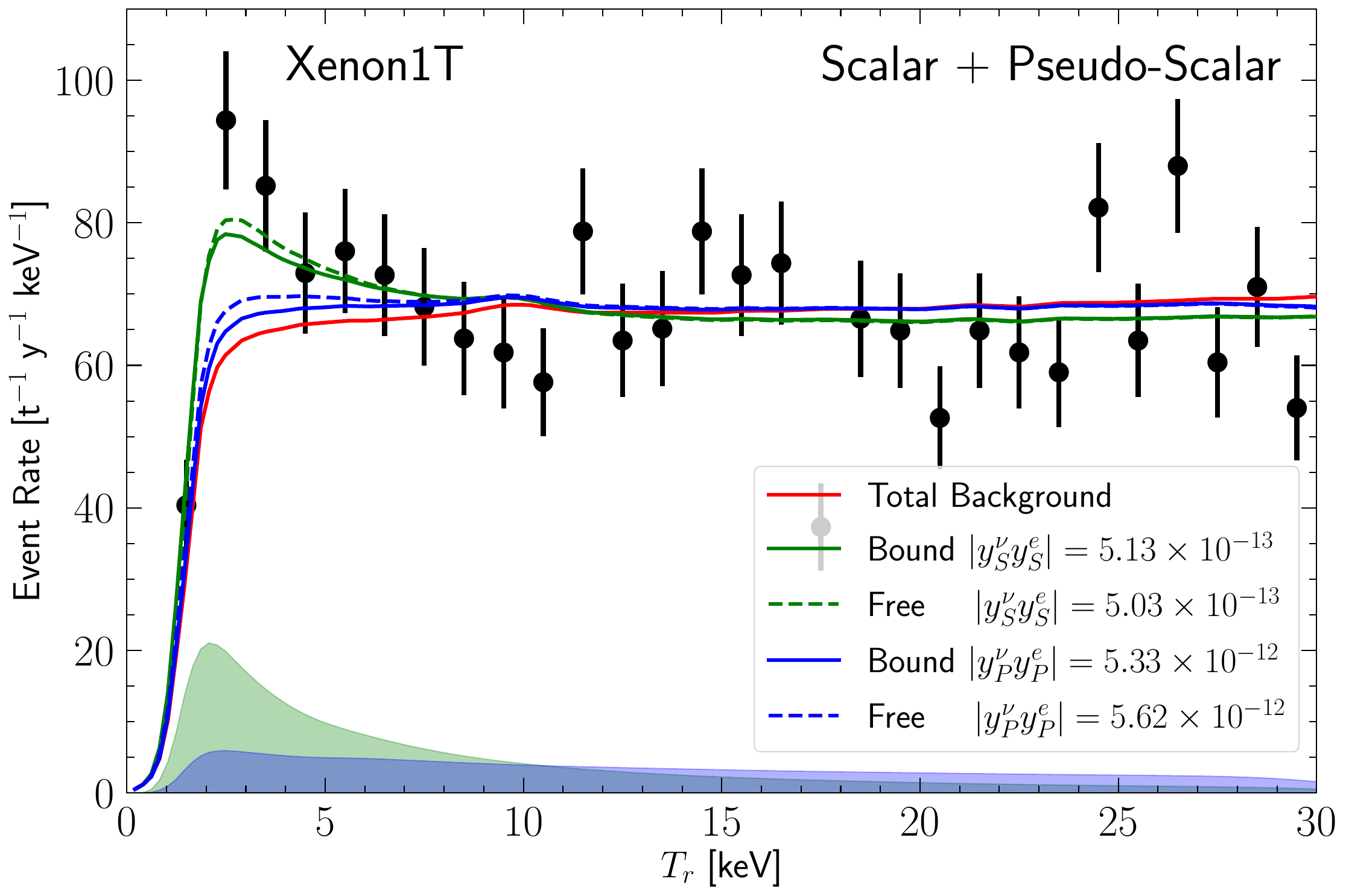}
\qquad
\includegraphics[width=8.4cm]{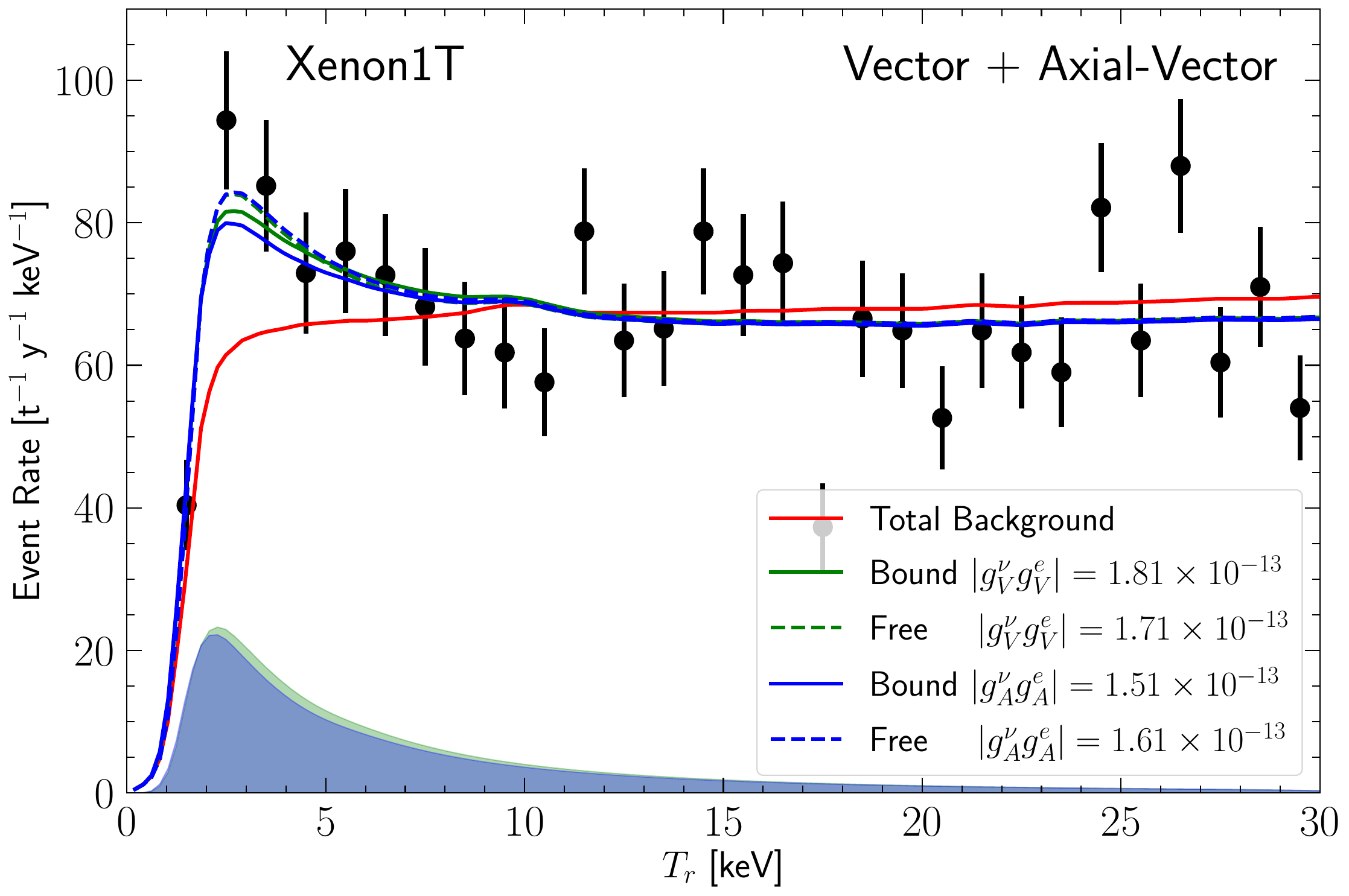}
\caption{Total events (black points) and background (red curve)
of XENON1T's data, we fit the data by solar neutrino-bound electron 
scattering with a sterile neutrino in final state, in addition
to both scalar (green) and pseudo scalar(blue) mediators. All
three parameters $m_s, m_\phi, |y_e y_p|$ are using the best-fitting values.
}
\label{Xenon1T_Events_mphi10}
\end{figure}

The XENON1T Collaboration has recently observed 
an excess in the electron recoil spectrum around $(2 \sim
3)$\,keV \cite{Aprile:2020tmw}. The experimental data is
shown as black points in \gfig{Xenon1T_Events_mphi10}
while the total background from radioactive materials
present in the detector and solar neutrino scattering
is shown as the red curve for
comparison. Note that the background is roughly flat and
hence can not explain the excess at low energy. It is
possible for this excess to arise from some new physics
and the data points can be used to constrain the
possible new physics model parameters.

In this paper, we give a more detailed evaluation of
the constraint on the solar neutrino scattering with
electrons into a massive sterile neutrino \cite{Ge:2020jfn}.
The atomic effects are also included to give a realistic result.
We use the following $\chi^2$ function,
\begin{eqnarray}
  \chi^2 
\equiv
  \sum_{i}^{\rm bins} 
  \left( 
    \frac{
      N^{\rm data}_i
    -  (1 + c) N^{\rm bkg}_i
    - N^{\rm NP}_i
    }{\sqrt{N^{\rm data}_i}}
  \right)^2
+
  \left(\frac c \sigma_c \right)^2,
  \label{eq:chi2_function}
\end{eqnarray}
to evaluate the best-fit value and sensitivities.
The $\chi^2$ function contains two major contributions:
the first term from the data bins and the second is
nuisance parameter $c$ for the background normalization
with uncertainty $\sigma_{c} = 2.6\%$ \cite{Aprile:2020tmw}.
In each bin, $N^{\rm data}_i$ is the data point,
$N^{\rm bg}_i$ the expected background,
and $N^{\rm NP}_i$ the expected events due to new
physics (NP) contribution. For the light mediator case
considered in this paper, $N^{\rm NP}_i$ is a function of
the mediator mass ($m_\phi$ or $m_{Z'}$), the sterile
neutrino mass ($m_s$), and couplings
($|y^\nu_{S,P} y^e_{S,P}|$ or $|g^\nu_{V,A} g^e_{V,A}|$).

\begin{figure}[t]
\centering
\includegraphics[width=8cm]{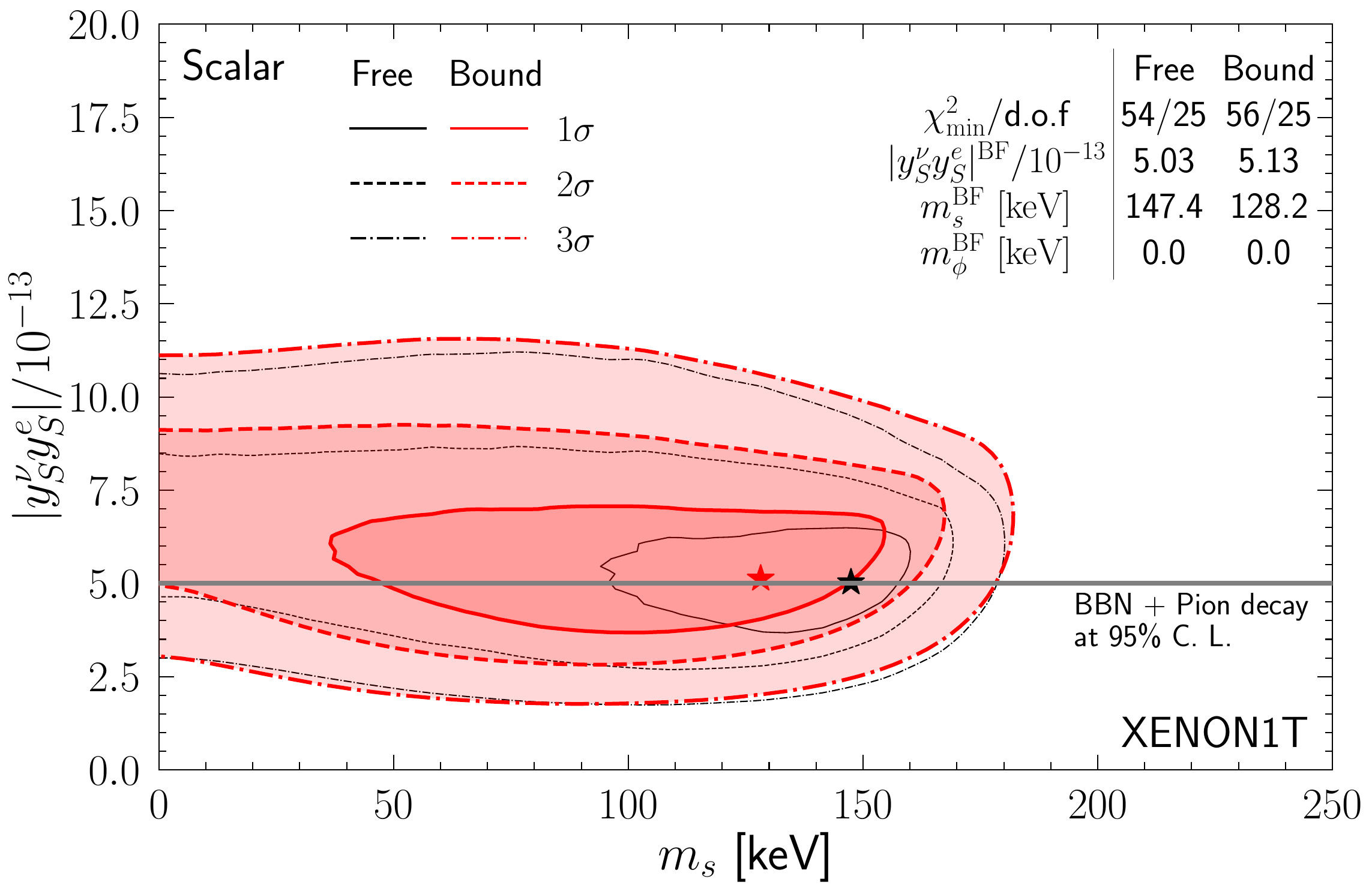}
\includegraphics[width=8cm]{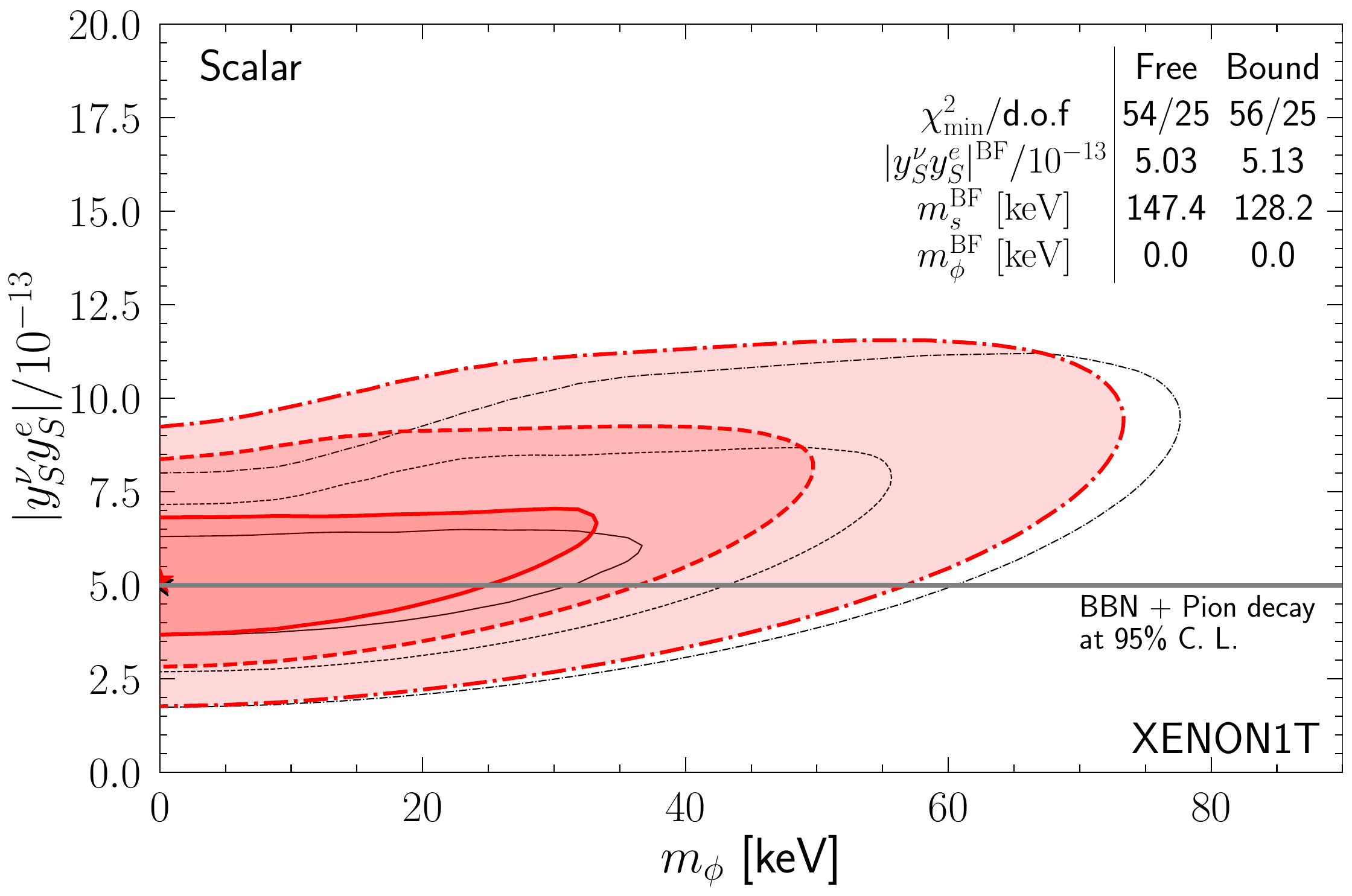}
\\
\includegraphics[width=8cm]{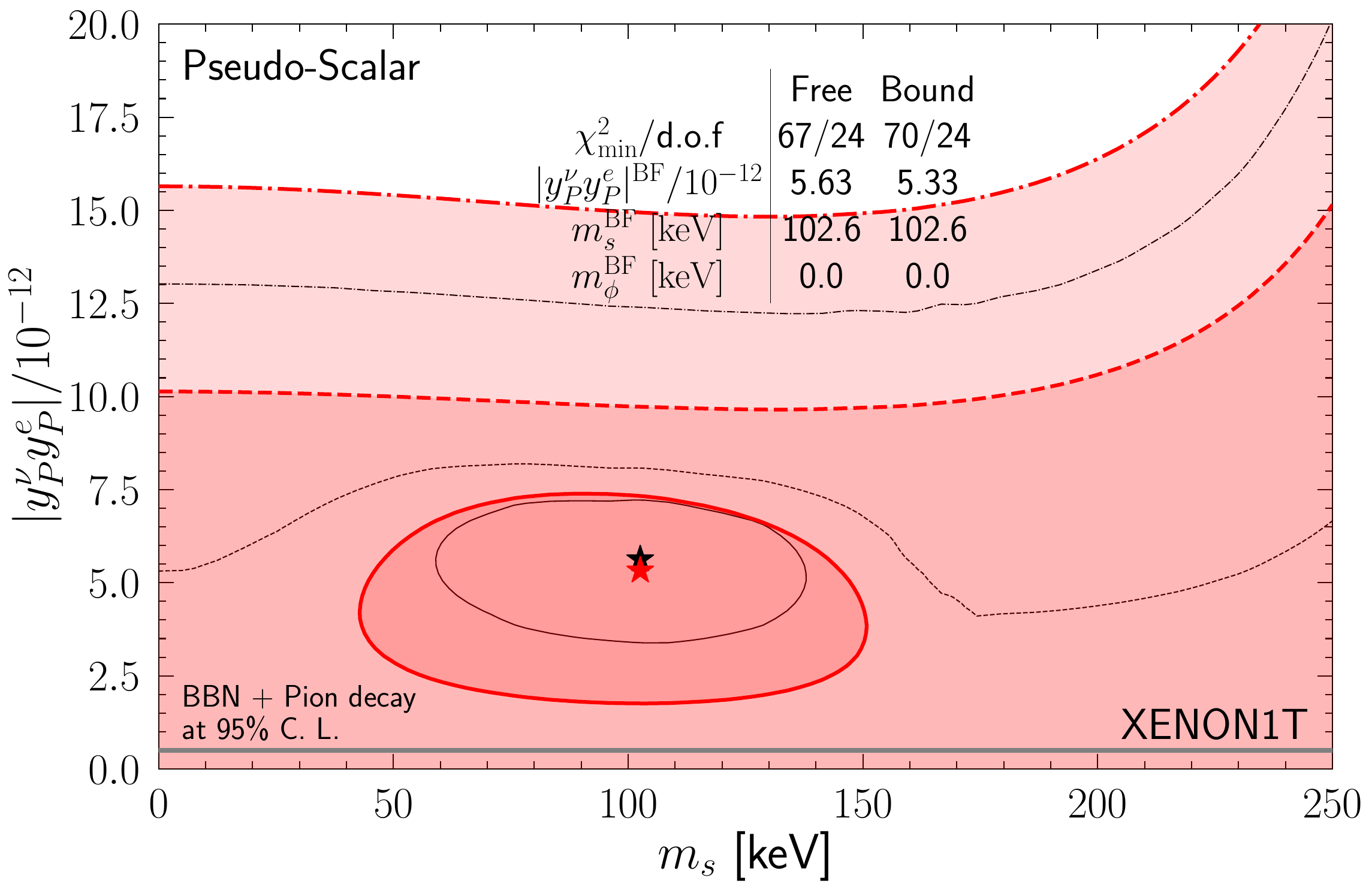}
\includegraphics[width=8cm]{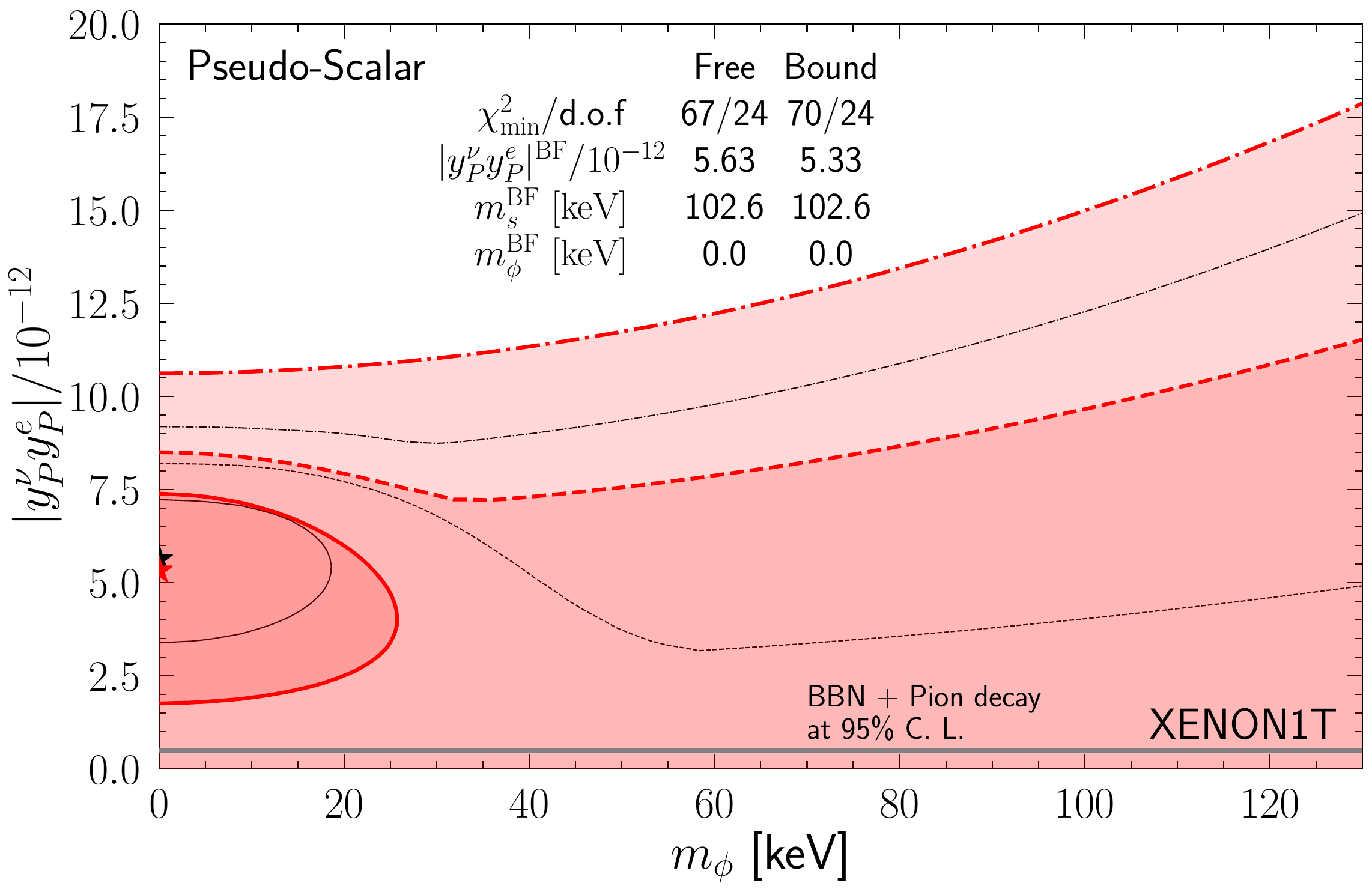}
\caption{
The XENON1T constraints for scalar (up), pseudo-scalar (bottom) interaction 
couplings $|y_{S,P}^\nu y_{S,P}^e|$ versus $m_s$ (left) 
and $m_\phi$ (right) at $1 \sigma$, $2 \sigma$,
and $3\sigma$ C.L. The black curves are for the free electron
approximation and red ones for the bound electron scenario whose
best-fit points are indicated with
black (free) and red (bound) stars, respectively.
The mediator (sterile neutrino) mass $m_\phi$ ($m_s$) is
marginalized over $0 \, \mbox{keV} < m_\phi < 175$\,keV  
($0 \, \mbox{keV} < m_s < 250$\,keV).
}
\label{Regionplot_scalar}
\end{figure}

On the theoretical side, the event rates are calculated by
the convolution of the differential cross-section defined in
\geqn{totalenergyspectrum} and the solar neutrino flux. We use the 
$pp$-neutrinos since they  have the largest flux at low energy 
\cite{Vinyoles:2016djt}. The detector's energy resolution 
is also taken into account by a Gaussian parametrization 
with variance given by $\sigma_{T_r} = a / \sqrt{T_r/{\rm keV}} + b$
with $a = 31.71 \pm 0.65\,$keV and $b = 0.15 \pm 0.02\,$keV
\cite{Aprile:2020yad}. In addition, the detector efficiency at 
a given recoil energy $T_r$ is extracted from 
\cite{Aprile:2020tmw}. \gfig{Regionplot_scalar} shows
the best-fit values and exclusion curves obtain by fitting
the predicted event numbers $N^{\rm NP}_i$ with data using
\geqn{eq:chi2_function}. The left and right panels are
obtained by fixing two parameters, $|y^\nu_{S,P} y^e_{S,P}|$ and
$m_\phi$ ($m_s$), and minimizing the $\chi^2$ function over
the remaining $m_s$ ($m_\phi$).

For comparison, each panel shows both free (black curves)
and bound (red areas) electron scenarios.
The bound-to-free ratio $R = 0.5$ for the scalar
mediator in \gfig{fig:total_diff_ratio} seems be quite different
from 1 and can have significant consequence. However, it is
mainly due to the broadening effect of the electron motion
inside atom. With detector smearing effect, the sharp peak
for the free scattering case would also become broader and
lower to more closely resemble the bound case. Consequently,
the coupling best-fit value almost does not change.
Take the scalar interaction as an example, the coupling
best fit marginally shifts from the free electron case
($|y^e_S y^\nu_S| = 5.03 \times 10^{-13} $) to the bound one,
$|y^e_S y^\nu_S| = 5.13 \times 10^{-13}$.
Similarly, the coupling strength changes from
$5.63 \times 10^{-12}$ to $5.33 \times 10^{-13}$
for pseudo-scalar mediator. 
The background-only hypothesis ($|y^e_{S,P} y^\nu_{S,P}| = 0$)
is excluded by more than $3\sigma$ C.L. for 
the scalar interaction but is still inside the 
$2\sigma$ region for the pseudo-scalar case. 
This is because the 
pseudo-scalar interaction has a smooth total 
differential cross section (blue curves in the left panel
of \gfig{fig:total_diff_ratio})
and is not as different from the background as the scalar one.

In addition to the DM direct detection,
the coupling with electron $y^e_{S,P}$
also receives constraints from other
experiments and astrophysical observations.
The torsion balance experiments impose a constraint 
$|y^e_{S,P}| < 10^{-23}$ at 95\% C.L. for the mediator 
mass below $4\times 10^{-6}$\,eV 
\cite{Adelberger:2009zz}, which is 10 orders smaller than 
the current DM direct detection constraints of $\mathcal{O}(10^{-13}$).
In other words, an ultra-light mediator
cannot explain the XENON1T excess. The mediator
mass up to $\mathcal O(0.01)$\,eV is also 
excluded by the Red Giant (RG) 
and Horizontal Branch (HB) stellar cooling
constraint $|y^e_{S,P}| < 10^{-16}$ \cite{Hardy:2016kme} 
at 95\% C.L. that 
is 3 orders smaller than the DM direct detection ones. 
The Big Bang nucleosynthesis (BBN) 
constrains the mediator mass above $10^{-2}$\,eV.
The $\phi$ production in the early 
Universe increases the relativistic degrees of 
freedom and accelerates the universe expansion.
A faster expansion reduces the
deuterium abundance. The $\phi$ production 
can only be evaded 
if $|y^e_{S,P}| < 5\times10^{-10}$ at 95\% C.L. or $m_\phi > 1$\, MeV \cite{Babu:2019iml}.
For the coupling with neutrino,  
the meson decay experiments require $|y_{S,P}^\nu| 
\lesssim 10^{-3}$ if $m_s < m_\pi$ \cite{Pasquini:2015fjv, Berryman:2018ogk, Dror:2020fbh, deGouvea:2019qaz}. 
Altogether,
the combined constraint is $|y^e_{S,P} y^\nu_{S,P}| \lesssim 5
\times 10^{-13}$ in the mass region $m_\phi > 
10^{-2}$\,eV (the gray lines 
in \gfig{Regionplot_scalar}).  
The best-fit value of 
the scalar case is allowed while 
the pseudo-scalar one is in tension.

\begin{figure}[t]
\centering
\includegraphics[width=8cm]{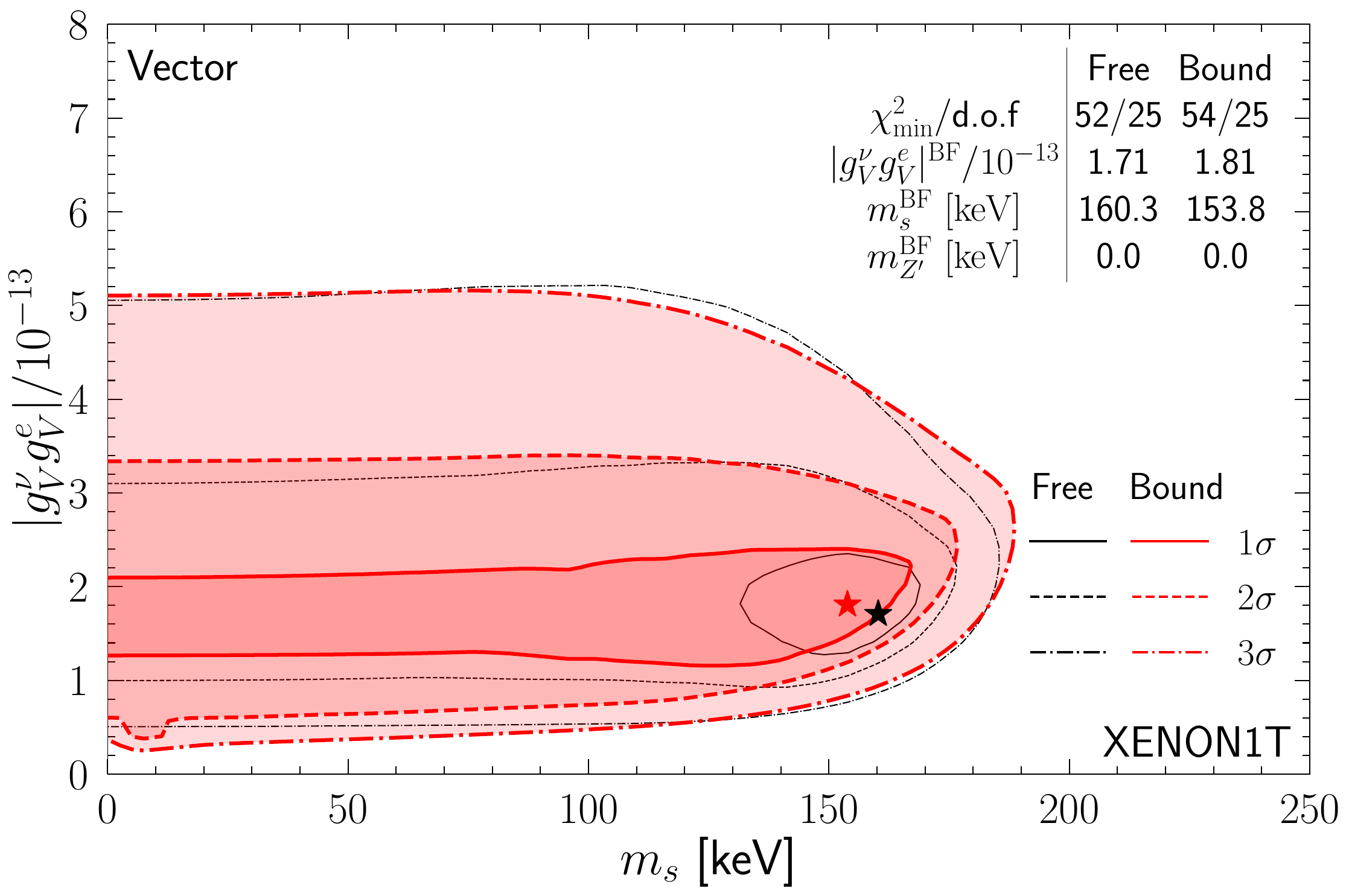}
\includegraphics[width=8cm]{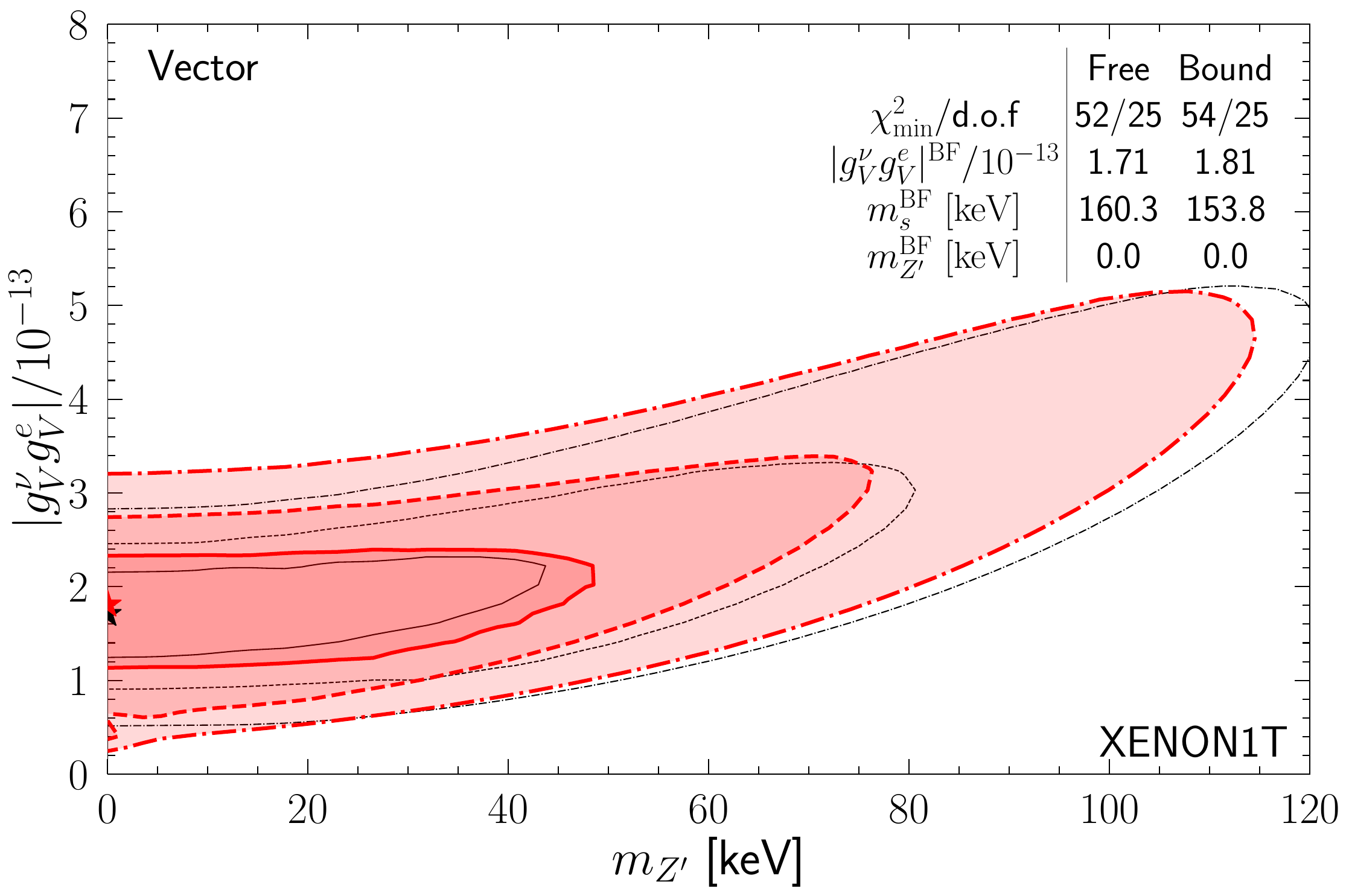}
\\
\includegraphics[width=8cm]{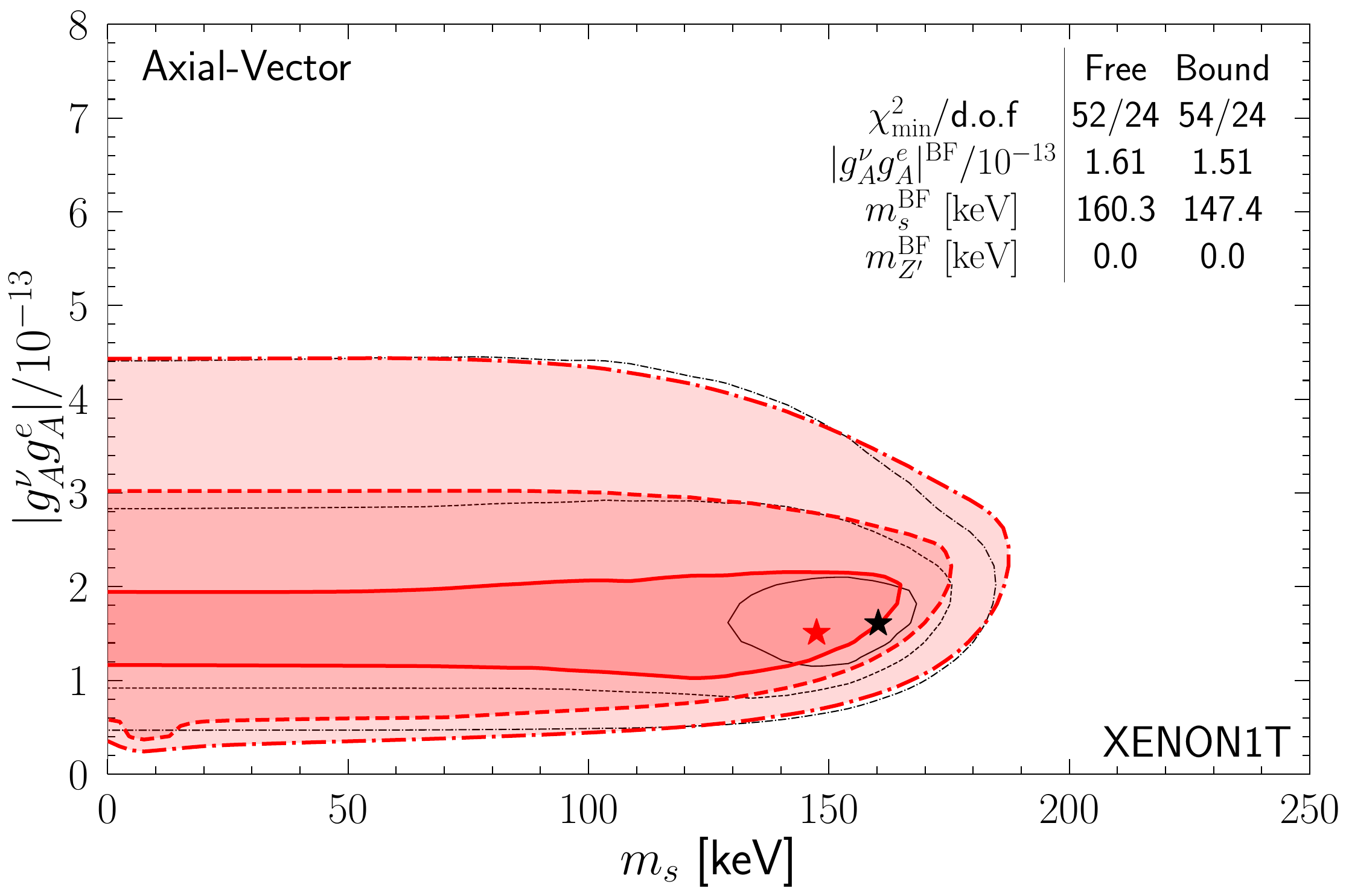}
\includegraphics[width=8cm]{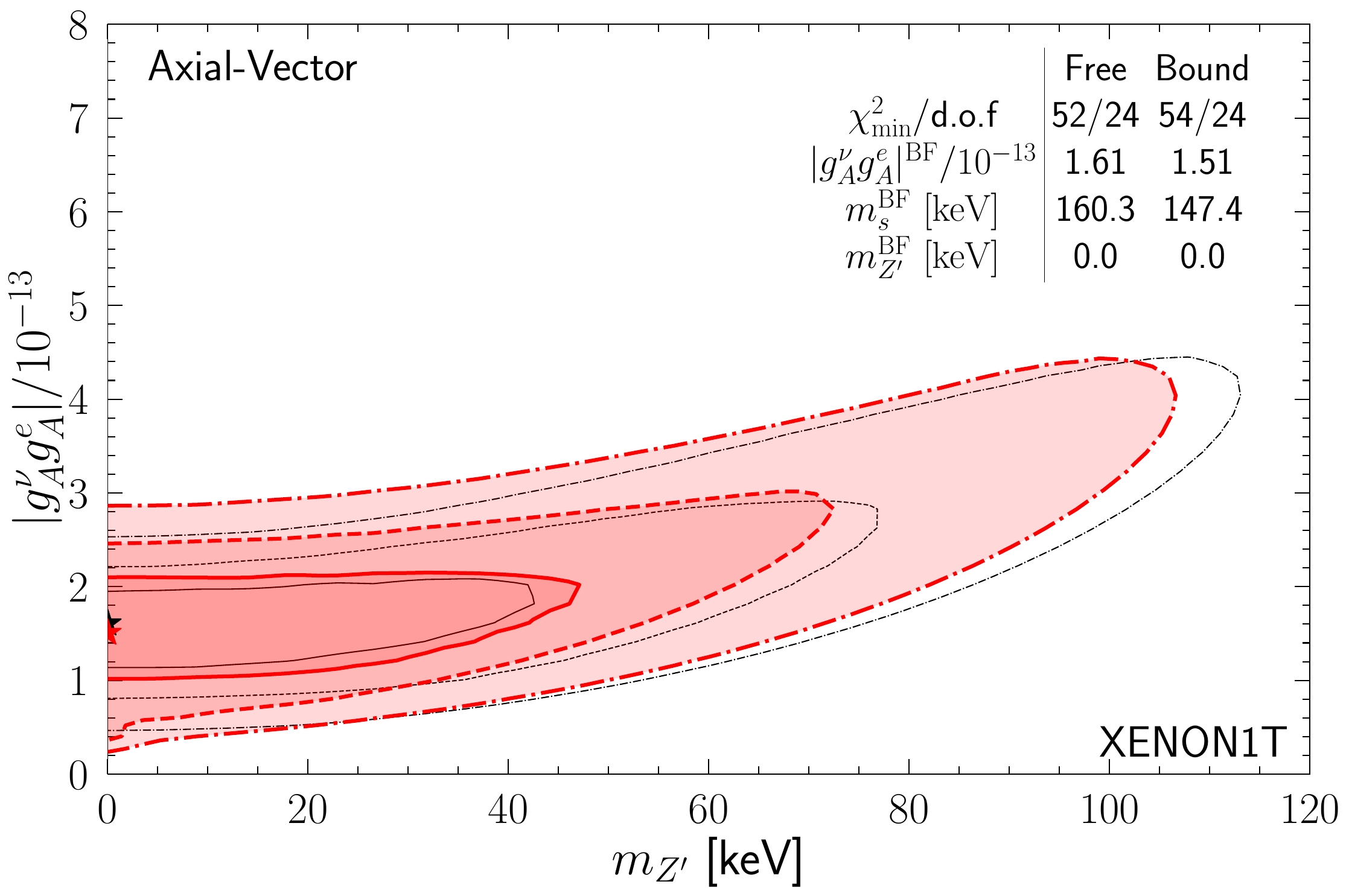}
\caption{
The XENON1T constraints for vector (up), axial-vector (bottom) interaction 
couplings $|g_{V,A}^\nu g_{V,A}^e|$ versus $m_s$ (left) 
and $m_\phi$ (right)
at $1,2$ and $3\sigma$ C.L. The same as before, 
black curves are for free electron approximation 
and red ones for bound electron scenario. Their best-fit values
are indicated with black (free) and red (bound) stars.
The mediator (sterile neutrino) mass $m_{Z'}$ ($m_s$) is
marginalized over $0 \, \mbox{keV} < m_{Z'} < 175$\,keV  
($0 \, \mbox{keV} < m_s < 250$\,keV).}
\label{Regionplot-vector}
\end{figure}

The best-fit value of sterile neutrino mass almost does not change for the scalar interaction, the best fit shifts slightly from $m_s = 147.4$\,keV to $128.2$\,keV after implementing
atomic effects. For the pseudo-scalar interaction, 
the best fit remains almost the same at $m_s = 102.6$\,keV. 
In other words, the dependence on the sterile neutrino mass is
not sensitive to the atomic effects. Similar feature also
applies for the light mediator mass whose best-fit value
is 0\,keV in all situations. 
This is because the propagator contribute a 
factor $\propto 1/(|\textbf{q}|^2 - \Delta E^2_{nl} + 
m_\phi^2)^2$ in the differential cross section
which typically grows with decreasing $m_\phi$ for given
$|{\bf q}|^2$. In other words, a smaller mediator mass leads
to a higher peak at lower recoil energy which is preferred 
by the XENON1T data.
However, the light mediator mass cannot really be zero
due to various constraints \cite{Babu:2019iml,Adelberger:2009zz}. 
To be on the safe side, we adopt $m_\phi = 10$\,keV 
which is still inside the $1 \sigma$ range when
predicting the event rates in \gsec{Section:Predictions}.
A mediator mass of 10\,keV avoids the constraints
mentioned before. This modification does not make 
big difference in the direct detection 
measurement. 
To be more concrete, the $\chi^2$ only increases by 1 and 
1.5 for the scalar and pseudo-scalar cases, respectively.

\gfig{Regionplot-vector} shows the results for vector and
axial-vector interactions. We can see that the basic features
are similar to the scalar and pseudo-scalar counterparts.
The coupling $|g_V^\nu g_V^e|$  
changes from $1.7 \times 10^{-13}$ 
to $1.8 \times 10^{-13}$ for vector
while $|g_A^\nu g_A^e|$ changes from 
$1.6 \times 10^{-13}$ to $1.5 \times 10^{-13}$ for axial
vector interactions.
This change can also be 
understood by the combination of bound-to-free cross section ratio and the detector smearing effect in a
similar way as the scalar/pseudo-scalar cases above.
Both vector and axial-vector interactions can 
exclude the background-only hypothesis by
more than $3\sigma$.
The best-fit value 
of sterile neutrino mass is around $150$\,keV
and the mediator mass still prefers a tiny value.
Similarly to the scalar and pseudo-scalar cases, we take
$m_{Z'} = 10$\,keV below when 
making projections for the future experiments.

The most stringent model-independent constraint on the 
coupling with electron, $|g^e_{V,A}| < 5\times 10^{-10}$, 
also comes from BBN  \cite{Ibe:2020dly}. 
In the sterile neutrino mass region $m_s < m_\pi$,
the leptonic pion decay imposes a bound $|g^\nu_{V,A}| < 0.014$
on the coupling with neutrino
\cite{Bakhti:2017jhm,Dror:2020fbh}. The combined result 
is $|g^e_{V,A}g^\nu_{V,A}| < 7\times 10^{-12}$. 
For both the vector and axial-vector interactions,
the best-fit points are well below this constraint.
Both cases can escape the constraints.
In fact, the bound is one order of magnitude higher than the
best-fit points and is not visible in \gfig{Regionplot-vector}

\section{Predictions for Future DM Experiments}
\label{Section:Predictions}

Although the XENON1T excess can be explained by new physics, 
the significance is not large enough and
there is still no definite conclusion
for the DM-electron interaction \cite{Aprile:2020tmw}.
Especially, the tritium
background is also  possible explanation
\cite{Aprile:2020tmw, PandaX-II:2020udv}.
More data is necessary to obtain a conclusive result.
In this section, we use the best-fit values
from the current XENON1T data to make prediction
for future experiments. 

We focus on three major DM direct detection experiments
with liquid Xenon target. First, the PandaX-4T experiment
\cite{Zhang:2018xdp} that has already started running
in 2021 is an upgrade of PandaX-II with a fiducial mass
of 2.8\,t and a factor of 10 improvement in the sensitivity.
Next is the XENONnT experiment \cite{Aprile:2020vtw} 
upgraded from XENON1T. XENONnT has a fiducial mass of
4\,t and can reduce its background by a factor of 6.
Finally, the LUX-Zeplin (LZ) experiment is a combination of two existing
experiments LUX \cite{Akerib:2016vxi} 
and Zeplin-III \cite{Akimov:2011tj}. Its fiducial mass
can reach 5.6\,t and also has very low background.
In addition, we make predictions with the same efficiency
of PandaX-II \cite{Tan:2016zwf} for PandaX-4T
and XENON1T \cite{Aprile:2020tmw} for XENONnT while for
LZ the Fig.\,3 of \cite{Akerib:2018lyp} is used.

\begin{figure}[t]
\centering
\includegraphics[width=8.5cm]{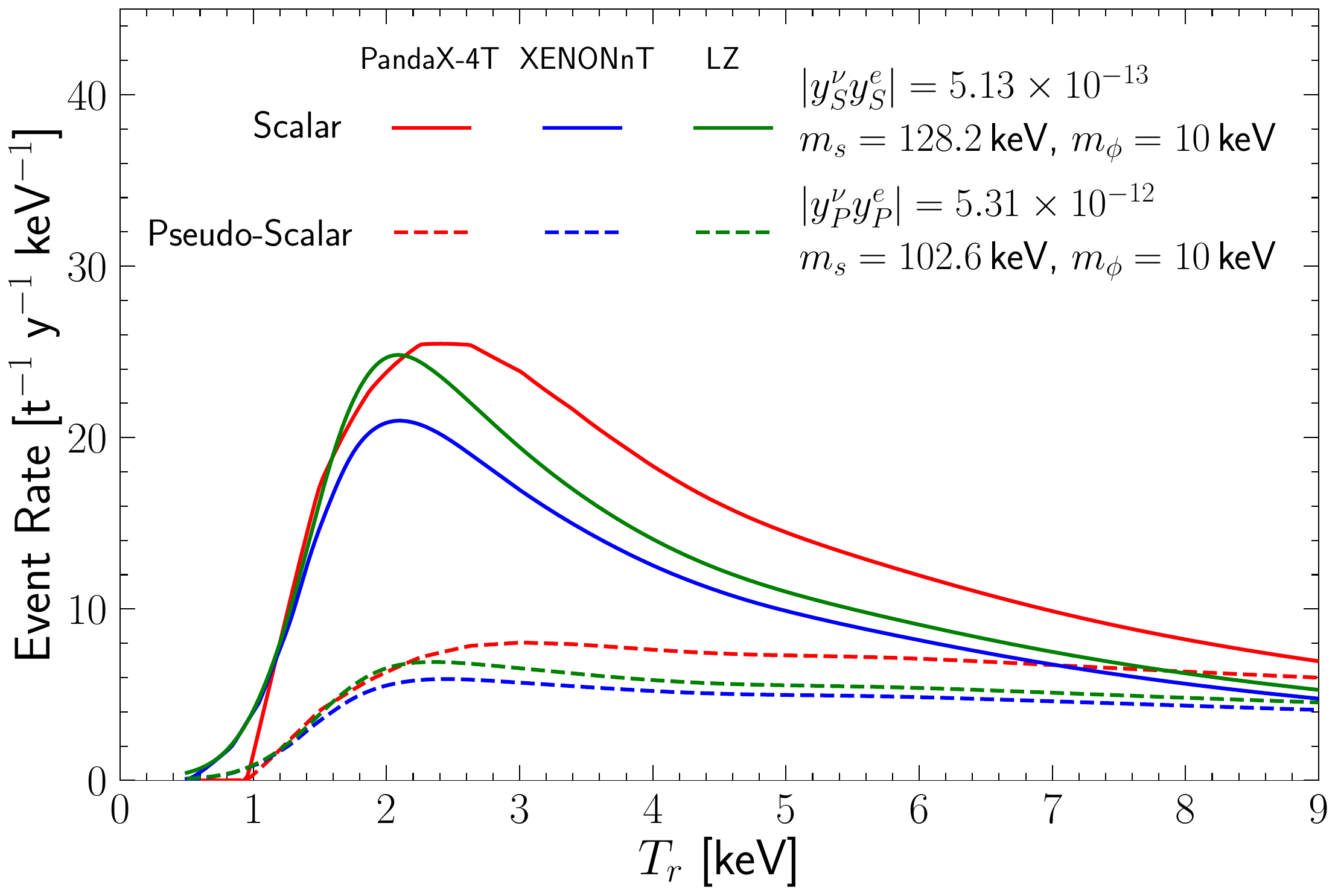}
\includegraphics[width=8.5cm]{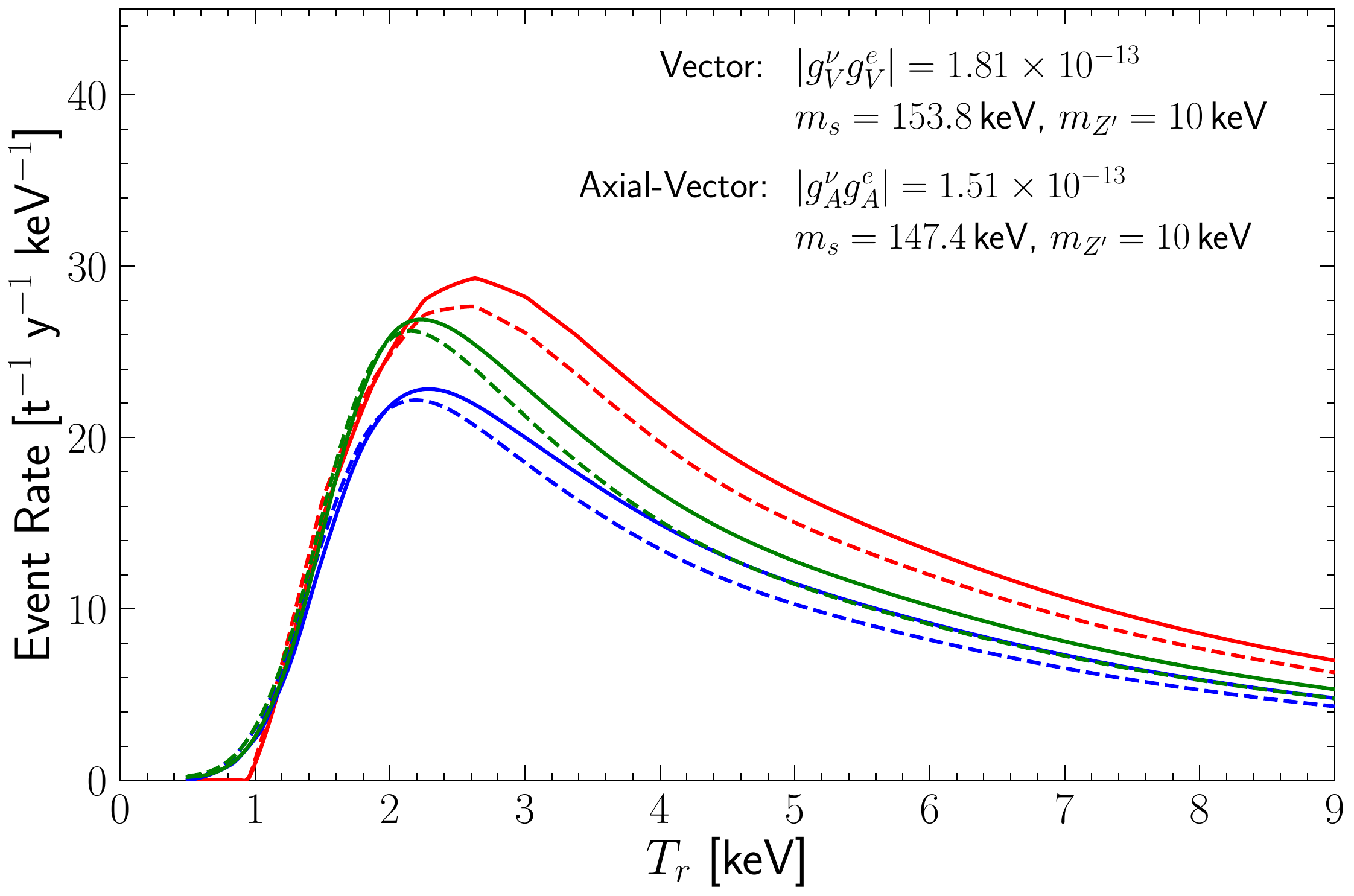}
\caption{
  Predictions for three future 
  experiments, PandaX-4T (red), XENONnT (blue),
  and LZ (green). 
  left plot is for scalar (solid)
  and pseudo-scalar (dashed) interactions
  while right plot for vector (solid)
  and axial-vector (dashed) interactions.}
\label{fig:prediction2}
\end{figure}
Fig.\ref{fig:prediction2} shows the predicted event rates
with atomic effects for future experiments
PandaX-4T (red), XENONnT (blue), and LZ (green). 
The left panel shows the results for scalar (solid)
and pseudo-scalar (dashed) interactions while the right
panel shows the vector (solid) and axial-vector (dashed) cases.
In these predictions, the couplings and the sterile neutrino
mass $m_s$ are assigned the best-fit values with the XENON1T data.
But the light mediator masses ($m_\phi$ and $m_{Z'}$) take a
universal value of 10\,keV. With atomic effects, the event rates 
for scalar, vector and axial-vector interactions have a more 
conspicuous low energy excess around $(2 \sim 3)$\,keV.
The event rate for the pseudo-scalar interaction gradually
increases towards low energy.

\begin{figure}[t]
\centering
\includegraphics[width=8.5cm]{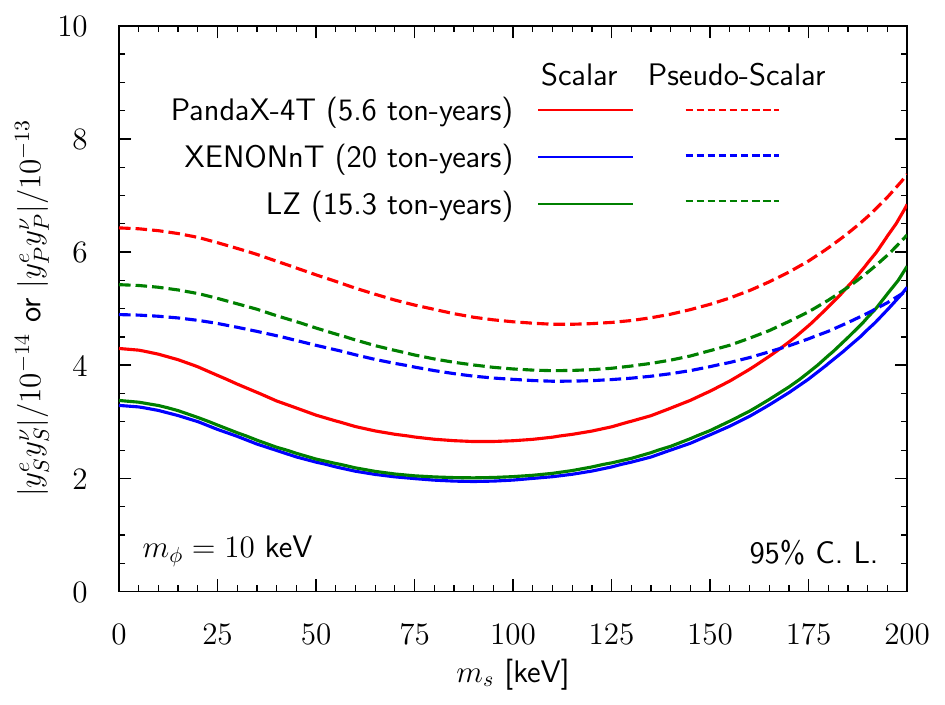}
\includegraphics[width=8.5cm]{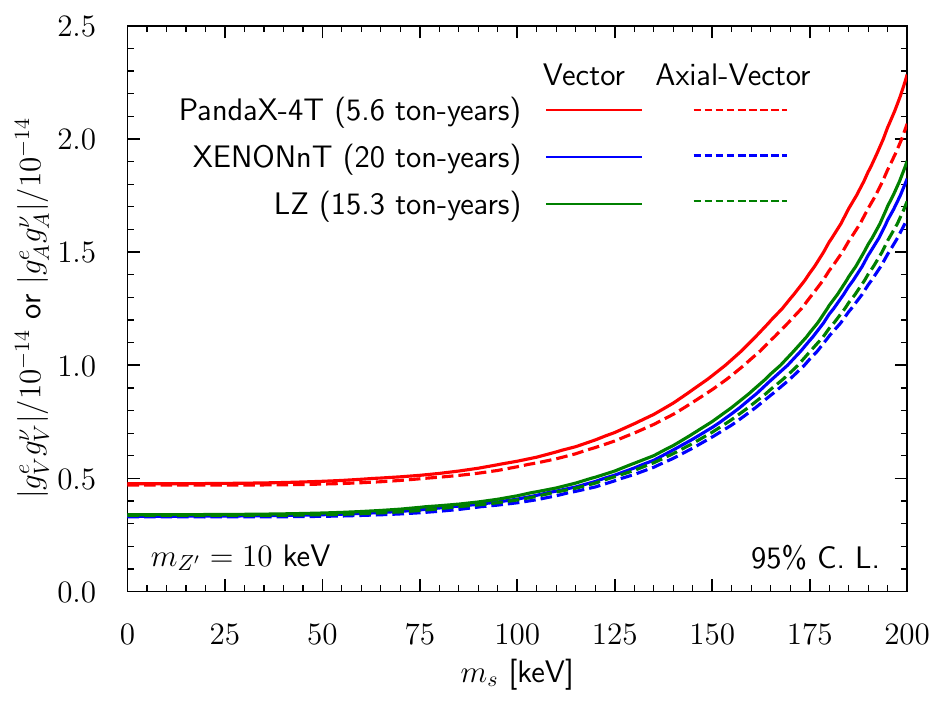}
\caption{
  The PandaX-4T (red), XENONnT (blue),
  and Lux Zeplin (green) projected 95\% C.L.
  exclusion curves as a function of the 
  sterile neutrino mass $m_s$.
  The left panel is for the scalar (solid) 
  and pseudo-scalar (dashed)
  mediators while the right panel is for 
  the vector (solid) and axial-vector
   (dashed) interactions. In all cases, the
   mediator mass is fixed at $m_{\phi,Z'} =
   10$\,keV.
  }
\label{fig:sensitivity_future}
\end{figure}

\gfig{fig:sensitivity_future} shows the 
expected sensitivity of the scalar, 
pseudo-scalar (left) and vector, 
axial-vector (right) couplings as
function of the sterile neutrino mass 
$m_s$, for the three future DM 
experiments PandaX-4T, Lux-Zeplin, and XENONnT. 
Their nominal exposure are 5.6, 15.3, and
20 ton-years, respectively. The results
represent a benchmark where the radioactivity
background is reduced to a negligible
level.
Only the irreducible solar neutrino background 
is considered in the analysis.
There is no new physics presenting in
the pseudo data. For each $m_s$, we first
fix $m_\phi, m_{Z'} = 10$\,keV to obtain a
$\Delta \chi^2$ 
as function of the coupling constants. 
From this one-dimensional $\Delta \chi^2$,
we obtain the 95\% C.L.
limit with $\Delta\chi^2 =
3.84$ on couplings
as shown in \gfig{fig:sensitivity_future}.

The sensitivity varies among
PandaX-4T, XENONnT, and Lux-Zeplin
due to different exposure time
of 2 years, 5 years, and 1000 
days, respectively. 
All experiments can probe the best-fit values
in \gfig{Regionplot_scalar}
and \gfig{Regionplot-vector}. 
They can also improve the
sensitivity by roughly one order of
magnitude. 
For example, the best-fit value of pseudo-scalar 
interaction is $5.6 \times 10^{-12}$ at 
XENON1T and the constraint becomes
$\mathcal{O}(10^{-13})$ in \gfig{fig:sensitivity_future}.
With this improvement, the future DM experiments
can confirm or falsify the pseudo-scalar
explanation to the XENON1T excess.

\section{Conclusions}
\label{sec:summary}

The XENON1T electron recoil excess stimulated various
explanations with new physics including solar
neutrino scattering with light mediators. Although
the atomic effects are commonly considered for those
dark matter scenarios, they are omitted in the solar
neutrino explanations. In the first part of this
work, we establish a systematic second quantization
formalism for both the initial-state bound and
final-state ionized electrons. This approach
introduces the atomic $K$-factor in a natural way
with general interactions for both neutrino and dark
matter scatterings.
The $K$-factor calculation includes
a summation over the final-state angular quantum
number $l'$ to infinity. 
In practice, one needs to truncate 
the summation at some maximum value $l'_{max}$
and this introduces some calculation error. 
A sub-percent precision in the atomic
$K$-factor requires at least $l'_{max} = 200$ 
for the electron recoil energy 
of $\mathcal O(10)$\,keV.

This formalism is then applied to the solar neutrino
scattering with bound electrons into a sterile 
neutrino through scalar, pseudo-scalar, vector, and 
axial-vector interactions. The atomic effects decrease the cross section by $1 \sim 5$ times to modify the recoil energy spectrum. Besides, 
the electron momentum distribution inside atom 
smears the momentum transfer of scattering. 
Consequently, the differential cross sections become smoother and broader than the free case.
We then use the bound electron scattering 
scenario to fit the XENON1T excess. 
The scalar, vector, and axial-vector 
interactions are preferred over the 
background-only hypothesis by more than 
$3\sigma$ with best-fit coupling constants 
of $\mathcal{O}(10^{-13})$. Meanwhile, the 
pseudo-scalar case is favored over
the background-only hypothesis by less
than $2\sigma$ with a coupling constant 
of $\mathcal{O}(10^{-12})$. In all cases, the 
best-fit value of sterile neutrino mass is around $100$\,keV.
Our results demonstrate for the
first time that the atomic effects can not 
be ignored when using solar neutrino 
scattering to explain the XENON1T
electron recoil excess. 

Finally, we project
the signal event rate and sensitivity at PandaX-4T, XENONnT, and LZ. 
These future experiments 
improve the sensitivity by roughly 
one order.
The current best-fit values and the 
$1 \sigma$ C.L. regions of the 
scalar, vector, and axial-vector 
mediators are marginally below the
constraints from astrophysics and pion 
decay. For the pseudo-scalar case, most 
regions of the parameter space are in
tension with the current constraint. 
With a factor of $10$
improvement in sensitivity, the solar
active-sterile neutrino conversion with
bound electrons via light mediators as an
explanation to the XENON1T excess can be
tested.

\section*{Acknowledgements}

The authors would like to thank Yi-Fan Chen, Timon Emken,
Xiao Xue, and Ning Zhou for useful discussions.
This work is supported by the Double First Class start-up fund
(WF220442604), the Shanghai Pujiang Program (20PJ1407800),
National Natural Science Foundation of China (No. 12090064),
and Chinese Academy of
Sciences Center for Excellence in Particle Physics (CCEPP).

\end{document}